\documentclass[amssymb,prb,twocolumn,showpacs,floatfix]{revtex4}
\usepackage{amsmath}
\usepackage{bbm}
\usepackage{graphicx}
\usepackage{verbatim}
\newcommand{\bea}{\begin{eqnarray}}
\newcommand{\eea}{\end{eqnarray}}
\newcommand{\be}{\begin{equation}}
\newcommand{\ee}{\end{equation}}
\begin{document}

\title{\bf Resonance fluorescence in driven quantum dots: electron and photon correlations}
\author{ Rafael S\'anchez $^{1,2}$, Gloria Platero $^{1}$ and Tobias Brandes $^{3}$}

\affiliation{ 1- Instituto de Ciencia de Materiales de Madrid
(CSIC), Cantoblanco, 28049 Madrid, Spain \\
2- D\`epartement de Physique Th\'eorique, Universit\'e de Gen\`eve, CH-1211 G\`eneve 4, Switzerland\\
3- Institut f\"ur Theoretische Physik, Technische Universit\"at Berlin, D-10623 Berlin, Germany.}

\begin{abstract}
We study the counting statistics for electrons and photons being emitted from a driven two level quantum dot. Our technique allows us to calculate their mutual correlations as well. We study different transport configurations by tuning the chemical potential of one of the leads to find that the electronic and photonic fluctuations can be externally manipulated by tuning the AC and transport parameters. We also propose special configurations where electron-photon correlation is maximal meaning that spontaneous emission of photons with a well defined energy is regulated by single electron tunneling. Interesting features are also obtained for energy dependent tunneling.  

\end{abstract}

\maketitle
\section{Introduction}
%The analogy between quantum dots   ({\it artificial atoms})    and real atoms 
%(not in vain they are also known as {\it artificial atoms}) 
%allow them to be considered as designed systems where to test quantum optical properties\cite{tobiasRep}. 

Quantum dots offer an ideal playground for testing coherent and quantum optical effects in an artificially designed solid-state environment \cite{tobiasRep}, with the additional benefit of having electronic transport \cite{JAUHO,FLENSBERG} 
% textbooks
as a `spectroscopy' by coupling to external fermionic reservoirs and counting the flow of single electrons. Complex behaviour emerges through electron-electron interactions and the interaction between electrons and other bosonic excitations such as photons \cite{gloria} or phonons  \cite{tobias}.

%Electron-boson interactions in these artificial atoms  are not only restricted to the coupling to radiation\cite{gloria} but they also occur due to the host material vibrations\cite{tobias}. Another essential difference between real atoms and quantum dots is the coupling of the quantum system to fermionic reservoirs, so electronic transport \cite{JAUHO} becomes important and can be controlled by means of gate voltages. 
%As a particular case of a translation of an optical problem to quantum dot systems, coherent population trapping has been predicted for driven three level quantum dots\cite{tobiasRenzoni} as well as for an undriven triple quantum dot configuration\cite{michaelis}.

The exchange of ideas between mesoscopic transport and quantum optics has turned out to be quite fruitful. For instance, thermal electron antibunching was observed experimentally by performing Hanbury Brown-Twiss-type experiments in mesoscopic conductors\cite{elHBT}. This fermionic antibunching has also been used 
%to purchase 
for regular photon sources in p-n junctions\cite{singleph} and quantum dots\cite{michler,benson}. Reversely, bosonic statistics can be studied in quantum conductors such as beam splitters\cite{gabelli}, nanoelectromechanical systems\cite{merlo} or quantum point contacts\cite{beenakkerSch1}, where photon antibunching was predicted\cite{beenakkerSch2}.  Another example is coherent population trapping and dark states in multilevel atoms\cite{cpt,gray}, originally proposed for driven three level quantum dots\cite{tobiasRenzoni} and then extended to triple quantum dots in a simple triangle configuration\cite{michaelis}.

Quantum transport also benefits from the adoption of theoretical tools that are well established in quantum optics. Specially relevant in the last years has been the development of noise\cite{BUETTIKER92,blanter} 
% two papers
and full counting statistics \cite{levitovLesovik,bagretsNazarov,franz} for electrons. Here, many of the relevant ideas and techniques  were in fact  originally developed in quantum optics in the context of counting single photons that are emitted from a single atomic source\cite{glauber,kelley,scully,cook,lenstra,cook80}. Recently, electronic counting statistics has become experimentally accessible for incoherent transport through quantum dot (QD) systems by the analysis of the time-resolved current flowing through a quantum point contact electrostatically coupled to them\cite{GustavssonPRL,GustavssonPRB,Gustavsson3,fujisawa}. However, the backaction of the quantum point contact on the QD destroys its internal coherence. Though traces of coherence have been measured in shot noise through stacks of double quantum dots\cite{barthold}, the access to higher order cummulants is still a challenging problem. 

Our aim in this work is to study the influence of electronic transport on the photonic emission statistics in a quantum dot system, and vice-versa.
Two-level systems give particularly interesting features both for optical and transport quantities: in optics, resonance fluorescence in two level atoms is the simplest case of a quantum photon source where photon antibunching occurs\cite{mandel}. In transport, quantum dots with two or more capacitively coupled levels show electronic bunching in dynamical channel blockade configuration\cite{belzig}. As will be shown, these properties can be studied in a two level quantum dot which is illuminated by a resonant ac electric field, where bosonic resonance fluorescence (due to phonon or photon mediated relaxation processes) is modified by electronic transport, and dynamical channel blockade depends on both coupling to a boson bath\cite{yoHaug} and the driving parameters\cite{PRL}.

%The complete knowledge of the statistics and, in concrete, the properties of the fluctuations of the number of particles emitted from a quantum system has been a topic of intense studies in quantum optics \cite{glauber,kelley,scully,cook,lenstra,cook80} and, in more recent years, in quantum transport\cite{levitovLesovik,bagretsNazarov,blanter,sigmund,ramonTobias,djuricSearch}. In particular,  interesting features like an anti-bunching of photons emitted from a closed two-level atom under a resonant field\cite{mandel}, or a bunching of electrons tunneling through interacting two-levels quantum dots (QD) connected to fermionic reservoirs \cite{belzig,djuricT,kiesslich03,kiesslich06,thielman,groth} have been reported. The electronic case appears to be specially important since measurements of higher order electron noise correlations have been recently realized\cite{GustavssonPRL,GustavssonPRB,Gustavsson3,fujisawa}.

We show how the {\em combined} statistics of Fermions and Bosons is a very sensitive
tool for extracting information from time-dependent, driven systems.
In particular, phonon emission has been measured by its influence on the electronic current in two level systems\cite{ramon}. We analyze the electron and photon noises and find that they can be tuned back and forth between sub- and super-Poissonian character by using the strength of an ac driving field or the bias voltage. % as a parameter.
%The application of an ac potential to a two level QD embedded into a phonon bath and  coupled to two biased leads drastically modifies both the electron and phonon statistics.

%{\em Technique--}
For this purpose, we develop a general method to  simultaneously extract the full counting statistics of single electron tunnelling  and (boson mediated) relaxation events, as well as their mutual correlations.

This paper is organized as follows: In Sec. \ref{model} we present our system and how to obtain the counting statistics of electrons and photons which are calculated in Secs. \ref{prf}, \ref{secdcb} and \ref{sechb} for different chemical potentials in the right lead. In Sec. \ref{secselect}, we present a special configuration where electron-boson correlation is maximal. The energy-dependent tunneling case is studied in section \ref{secleveldep}. Finally, Sec. \ref{conclus} presents our conclusions. Due to the length of some of the analytical results, we include some appendices with the coefficients that allow their calculation.

\section{Model and technique}
% MOVED FORM ABOVE
Our system consists of a two level quantum dot (QD) connected to two fermionic leads by tunnel barriers. % and coupled to a phonon bath.
The Coulomb repulsion inside the QD is assumed to be so large that only single occupation is allowed ({\it Coulomb blockade} regime). The lattice vibrations induce, at low temperatures, inelastic transitions from the upper to the lower state.
In analogy to resonance fluorescence in quantum optics,
a time-dependent ac field with a frequency $\omega$  drives the transition between the two levels $\varepsilon_1$, $\varepsilon_2$ close to resonance,
$\Delta_\omega=\varepsilon_2-\varepsilon_1-\omega\approx 0$ (we will consider $e=\hbar=1$ through the whole text), which allows us to assume the rotating wave approximation.
%\cite{carmichael}
Thus, the electron in the QD is coherently delocalized between both levels performing {\it photon-assisted Rabi oscillations}\cite{gloria,pump}. For simplicity, we consider spinless electrons. Studies of spin currents in a quantum dot coupled to a quantized driving or an AC magnetic field can be found in Refs. ~\cite{djuricSearch} and ~\cite{dong}, respectively. 

\label{model}
This system is modelled by the Hamiltonian:
\begin{eqnarray}\label{hamRF}
\hat H(t)&=&\sum_{i}\varepsilon_i\hat d_i^\dagger\hat d_i+\frac{\Omega}{2}\left(e^{-i\omega t}\hat d_2^\dagger\hat d_1+{\rm H.c.}\right)\nonumber\\
&&+\sum_{k\sigma}\varepsilon_{k\alpha}\hat c_{k\alpha}^\dagger\hat c_{k\alpha}+\sum_{k\alpha i}V_{\alpha i}\left(c_{k\alpha}^\dagger\hat d_i+ {\rm H.c.}\right)\\
&&+\sum_{q,\nu}\omega_q\hat a_{q\nu}^\dagger\hat a_{q\nu}+\sum_{ij,q\nu}\lambda_{q\nu}^{ij}\hat d_i^\dagger\hat d_j\left(\hat a_{-q\nu}+\hat a_{q\nu}^\dagger\right),\nonumber
\end{eqnarray}
where $\hat a_{q\nu}$, $\hat c_{k\alpha}$ and $\hat d_i$ are
annihilation operators of bosons (with momentum $q$ and polarization $\nu$) and electrons in the lead $\alpha$  and in the level $i$ of the QD, respectively, and $\Omega$ is the Rabi frequency, which is proportional to the intensity of the ac field.

The terms proportional to $V_{\alpha i}$ and $\lambda_{q\nu}^{ij}$ in Eq. (\ref{hamRF}) give the coupling of the electrons in the QD to the fermionic leads and their 
%{\it dipolar} ????????????????
interaction with the bosonic bath, respectively. 
In the following, we assume a basis of quantum dot levels where diagonal matrix elements of $\lambda_{q\nu}^{ij}$ play a minor role and we set $\lambda_{q\nu}^{ii}=0$. For coupling to photons, this would correspond to a usual dipole interaction with the electromagnetic field. For phonons, this is justified if we are mainly interested in weak coupling and relaxation processes by spontaneously emitted bosons in the relaxation from the upper to the lower level. The electron-boson coupling term, in the rotating wave approximation, can then be written as $\sum_{q\nu}\lambda_{q\nu}\left(\hat d_2^\dagger\hat d_1\hat a_{q\nu}+{\rm H.c.}\right)$. 
%%%%%% NEW
For the sake of illustrating our results with concrete physical processes we will  refer to {\em photon emission} in the following, i.e. the bosonic bath corresponds to the photon vacuum without additional coupling to phonons.  
%%%%%% NEW

% We point out, however, that our study is valid for the spontaneous emission of phonons as well
%. Depending on energy scales and device parameters, either of these two couplings might be the dominant one.
Finally, the coupling to the fermionic and bosonic baths terms are responsible for the incoherent dynamics and they can be considered apart in the derivation of the master equation for the reduced density matrix.

%%%%%%%%%%%%%%%%%%%
Considering the quantum-jump approach\cite{plenio,gurvitzPrager} to electronic transport and boson emission events, the equation of motion of the reduced quantum dot 
density matrix can be written as
%Defining ${\cal L}_{e(p)}$ as the Liouvillian superoperator describing the incoherent electron tunnelling and phonon emission events, we can write the master equation:
\begin{equation}
\dot\rho (t)={\cal L}_0(t)\rho (t)+{\cal L}_e(t)\rho (t)+{\cal L}_p(t)\rho (t),
\end{equation}
where ${\cal L}_e$ and ${\cal L}_p$ are the Liouvillian {\it jump} superoperators responsible for the incoherent events: electron tunneling from the system to the collector and relaxation by spontaneous photon emission. Thus, we can consider a density matrix resolved in the number of accumulated electrons in the collector, $n_e$, and the number of emitted photons, $n_p$:
\begin{equation}
\rho (t)=\sum_{n_e,n_p} \rho^{(n_e,n_p)}(t),
\end{equation}
where $\rho^{(n_e,n_{p})}(t)$ gives the probability that, during a certain  time interval $t$, $n_{e}$ electrons have tunneled out of a given electron-photon system and $n_{p}$ photons have been emitted.

This allows us to define the generating function\cite{lenstra}:
\begin{equation}\label{cgf}
G(t,s_e,s_{p})=\sum_{n_e,n_{p}} s_e^{n_e} s_{p}^{n_{p}}\rho^{(n_e,n_{p})}(t),
\end{equation}
by introducing the electron (photon) counting variables, $s_{e(p)}$.
The derivatives of $G(t,s_e,s_{p})$ with respect to the counting variables give the correlations:
\begin{equation}\label{corr}
\frac{\partial^{\ell+m} trG(t,1,1)}{\partial s_e^\ell\partial s_{p}^{m}}=\left\langle \prod_{i=1}^{\ell}\prod_{j=1}^{m}(n_e-i+1)(n_{p}-j+1)\right\rangle.
\end{equation}
up to any order.
%Thus, we are able to obtain the mean  %count:
%number $\langle n_\alpha\rangle$, the variance $\sigma_\alpha^2=\langle n_\alpha^2\rangle -\langle n_\alpha\rangle^2$ (which give the $\alpha=e,p$ current and noise, respectively), or define the correlation between the electron and phonon counts, $\langle n_{e}n_{p}\rangle$.

%For this purpose, it is needed to
One can derive the equations of motion for the generating function as previously done for the density matrix:
\begin{equation}\label{eomgf}
\dot G(t,s_e,s_{p})=M(s_e,s_{p})G(t,s_e,s_{p}),
\end{equation}
that generalizes %obtained from
the master equation, $\dot\rho(t)=M(1,1)\rho(t)$.
The jump superoperators affect only the diagonal elements of the generating function in the same way that rate equations involve only the occupation probabilities --given by the diagonal elements of the density matrix. The electronic one acts as:
%Considering only the terms that change the number of electrons in the collector, we obtain for the matrix elements of the generating function:
\be
\left({\cal L}_e(t)G(t,s_e,s_p)\right)_{mm}=\sum_{k}\left(s_e\Gamma_{mk}^{+}+s_e^{-1}\Gamma_{mk}^{-}\right)G_{kk}(t,s_e,s_p),
\ee
where $\Gamma_{mk}^{\pm}$ is the rate for processes that increase/decrease the number of electrons in the collector by a transition from state $|k\rangle$ to state $|m\rangle$. 
%Only diagonal elements contribute to transport. 
The same can be done for photons, with the difference that the number of detected photons can only increase. Then, one only has to introduce the corresponding counting variables in those terms corresponding to the tunneling of an electron to the collector lead, and the emission of a photon. The relevant elements of the density matrix can be written as a vector, $\rho=(\rho_{00},\rho_{11},\rho_{12},\rho_{21},\rho_{22})^T$, where $\rho_{00}$ gives the occupation of the empty state, $\rho_{11}$ and $\rho_{22}$ correspond to the ground and excited electronic states, respectively, and $\rho_{12}$ and $\rho_{21}=\rho_{12}^*$ are the coherences. Then, for the case where the tunneling barriers are equal for both energy levels, i.e. $V_{\alpha1}=V_{\alpha2}$, the equation of motion of the generating function (\ref{eomgf}) is described, in the Born-Markov approximation, %\cite{blum},
by the matrix\cite{blum,carmichael}:
\begin{widetext}
\begin{equation}\label{MLR}
\displaystyle
M(s_e,s_{p})=\left(\begin{array}{ccccc}
-2\Gamma_{\rm L}-(f_1+f_2)\Gamma_{\rm R} & s_e\bar f_1\Gamma_{\rm R} & 0 & 0 & s_e\bar f_2\Gamma_{\rm R} \\
\Gamma_{\rm L}+s_e^{-1}f_1\Gamma_{\rm R} &  -\bar f_1\Gamma_{\rm R} & i\frac{\Omega}{2} & -i\frac{\Omega}{2} & s_{p}\gamma  \\
0 & i\frac{\Omega}{2} & \Lambda_{12}+i\Delta_{\omega} & 0 & -i\frac{\Omega}{2} \\
0 & -i\frac{\Omega}{2} & 0 &\Lambda_{12}-i\Delta_{\omega} & i\frac{\Omega}{2} \\
\Gamma_{\rm L}+s_e^{-1}f_2\Gamma_{\rm R} & 0 & -i\frac{\Omega}{2} & i\frac{\Omega}{2} & -\gamma-\bar f_2\Gamma_{\rm R}\\
\end{array}  \right),
\end{equation}
\end{widetext}
%where $\rho_{00}$ gives the occupation of the empty state, $\rho_{11}$ and $\rho_{22}$ correspond to the ground and excited electronic states, respectively, and $\rho_{12}$ and $\rho_{21}=\rho_{12}^*$ are the coherences, 
where, by further considering that the density of states in the leads is rather constant so $d_\alpha(\omega_{mn})=d_\alpha$, the tunneling rates through the lelft(right) lead are equal to $\Gamma_{{\rm L(R)}}=2\pi d_{\rm L(R)}|V_{\rm L(R)}|^2$. We will consider a high bias configuration where the chemical potential of the left lead is well above the energy levels and the transitions between the right lead (with a chemical potential $\mu$) and the state $i$ in the QD are weighted by the Fermi distribution functions $f_i=f(\varepsilon_i-\mu)=\left(1+e^{(\varepsilon_i-\mu)\beta}\right)^{-1}$ and $\bar{f}_i=1-f_i$. 
$\gamma$ 
%{\bf($=2\pi|\lambda_{\varepsilon_2-\varepsilon_1}|^2$  SUM OVER DELTA FUNCTION MISSING} ) 
is the spontaneous emission rate due to the coupling with the photon bath: $\gamma=2\pi\sum_\nu\int d^3qg(q)|\lambda_{q\nu}|^2\delta(|q|v-\varepsilon_2+\varepsilon_1)$, where $g(q)$ is the density of states\cite{carmichael}. 
%$\Gamma_{{\rm L(R)}}=2\pi|V_{\rm L(R)}|^2$ is the tunneling rate through the left(right) contact (considering $V_{\alpha1}=V_{\alpha2}$)
%and $f_i=f(\varepsilon_i-\mu)=\left(1+e^{(\varepsilon_i-\mu)\beta}\right)^{-1}$ and $\bar{f}_i=1-f_i$ are the Fermi distribution functions that weight the tunneling of electrons between the right lead (with a chemical potential $\mu$) and the state $i$ in the QD. 
The decoherence is given by $\Lambda_{12}=-\frac{1}{2}\left((\bar f_1+\bar f_2)\Gamma_{\rm R}+\gamma\right)$.
The Fermi energy of the left lead is considered high enough that no electrons can tunnel from the QD to the left lead.
All the parameters in these equations, except the sample-depending coupling to the photon bath, can be externally manipulated.

Taking the Laplace transform of the generating function, $\tilde G(z,s_e,s_{p})=(z-M)^{-1}\rho(0)$, where $\rho(0)$ is the initial state, the long-time behaviour is given by the residue for the pole near $z=0$.
% in the Laplace transform of the GF, $\tilde G(z,s_e,s_{p})=(z-M)^{-1}\rho(0)$, where $\rho(0)$ is the initial state.
From the Taylor expansion of the pole $z_0=\sum_{m,n>0}c_{mn}(s_e-1)^m(s_{p}-1)^n$, one can write %a generating function of the form:
$trG(t,s_e,s_{p})\sim {\textsl g}(s_e,s_{p})e^{z_0t}$ and obtain, from (\ref{corr}), the mean value, $\langle n_{e(p)}\rangle$, as well as the higher order cumulants, $\kappa_{e(p)}^{(i)}=\left\langle\left(n_{e(p)}-\langle n_{e(p)}\rangle\right)^i\right\rangle$:
% of the distribution function,
%asymptotic central moments of the distribution function from the coefficients $c_{mn}$:
\begin{subequations}
\begin{equation}
\label{k2}
\kappa_{e(p)}^{(2)}=
%\langle n_{e(p)}^2\rangle-\langle n_{e(p)}\rangle^2=
\frac{\partial^2 {\textsl g}(1,1)}{\partial s_{e(p)}^2}-\left(\frac{\partial {\textsl g}(1,1)}{\partial s_{e(p)}}\right)^2+(c_{10(01)}+2c_{20(02)})t
\end{equation}
\begin{eqnarray}
\label{k3}
\kappa_{e(p)}^{(3)}&=&\frac{\partial^3 {\textsl g}(1,1)}{\partial s_{e(p)}^3}-3\frac{\partial {\textsl g}(1,1)}{\partial s_{e(p)}}\frac{\partial^2 {\textsl g}(1,1)}{\partial s_{e(p)}^2}+2\left(\frac{\partial {\textsl g}(1,1)}{\partial s_{e(p)}}\right)^3\nonumber\\[-2.5mm]
&&\\[-1.4mm]
&&+\frac{\partial {\textsl g}(1,1)}{\partial s_{e(p)}}+(c_{10(01)}+6c_{20(02)}+6c_{30(03)})t,\nonumber
\end{eqnarray}
\end{subequations}
which give the variance and skewness of the probability distribution, respectively.
In the large time asymptotic limit,
all the information is included in the coefficients $c_{mn}$. Thus, we obtain the stationary current and the low frequency noise:
\bea
I_{e(p)}&=&\frac{d}{dt}\langle n_{e(p)}\rangle= c_{10(01)}\\
S_{e(p)}(0)&=&\frac{d}{dt}\kappa_{e(p)}^{(2)}= c_{10(01)}+2c_{20(02)},
\eea
respectively. Then,
the Fano factor is $F_{e(p)}=1+2c_{20(02)}/c_{10(01)}$ so that,
the sign of the second term in the right hand side defines the sub- ($F<1$) or super-($F>1$) Poissonian
 character of the noise. %%(whenever the Fano factor is smaller or greater than 1).

In the limit $\Gamma_{L(R)}\rightarrow 0$, the pure Resonance
Fluorescence case for the noise of the emitted photons, formally
equivalent to the expression for emitted photons in quantum
optics\cite{lenstra} is obtained:
\begin{equation}\label{frf}
F_{p}(\Gamma_i=0)=1-\frac{2\Omega^2(3\gamma^2-4\Delta_\omega^2)}{(\gamma^2+2\Omega^2+4\Delta_\omega^2)^2},
\end{equation}
yielding the famous sub-Poissonian noise result at resonance
($\Delta_\omega=0$). The detuning between the frequency of the AC field and the levels energy separation, $\Delta_\omega$, contributes to {\it restore} super-Poissonian statistics, as seen in Fig. \ref{phrf}. In the following, only the resonant case will
be considered unless the opposite were indicated.

Electron-photon correlations are obtained from:
\begin{equation}
\langle n_{e}n_{p}\rangle=\frac{\partial {\textsl g}(1,1)}{\partial s_{e}}c_{01}t+
\frac{\partial {\textsl g}(1,1)}{\partial s_{p}}c_{10}t+c_{11}t+c_{10}c_{01}t^2
\end{equation}
Then, by defining $\sigma_{ij}^2=\langle n_{i}n_{j}\rangle-\langle n_{i}\rangle\langle n_{j}\rangle$, where $\sigma_{ii}^2=\kappa_i^{(2)}$,
\begin{equation}
\sigma_{ep}^2=\frac{\partial^2 {\textsl g}(1,1)}{\partial s_{e}\partial s_p}+c_{11}t
\end{equation}
defines the variance between electronic and photonic events.
The long time behaviour is given by $\sigma_{ep}^2\sim c_{11}t$ so the correlation coefficient can then be defined as\cite{zajarov}:
\begin{equation}
r=\frac{\sigma_{ep}}{\sqrt{\sigma_{ee}^2\sigma_{pp}^2}}=\frac{c_{11}}{\sqrt{(c_{10}+2c_{20})(c_{01}+2c_{02})}}.
\end{equation}

Similarly to the electronic(photonic) correlations, where the sign of the second order cummulant, $c_{20(02)}$,
defined the sub- or super-Poissonian character of the noise, the sign of $c_{11}$
 gives the character of the electron-photon correlations. If $c_{11}>0$, the detection of
 a transmitted electron would involve the detection of a photon in a short lapse of time, while $c_{11}<0$
 involves distant events.

The electron-photon correlation coefficient is limited to $|r|<1$, having $r=1$ for the case
 where the number of detected electrons is proportional to the number of detected photons:
 $n_e\propto n_p$. $r=0$ means uncorrelated events. Note that independent events give $r=0$,
but the opposite is not necessarily true, as will be shown below.
 Analogously to the Fano factor for the second order cumulants,
 the deviation of the third cumulants from the Poissonian statistics
can be parametrized by the coefficient:
\begin{equation}
\eta_{e(p)}=\frac{1}{I_{e(p)}}\frac{d}{dt}\kappa_{e(p)}^{(3)}=1+6\frac{c_{20(02)}+c_{30(03)}}{c_{10(01)}}.
\end{equation}

In what follows, different configurations will be discussed
concerning the relative positions of the energy levels with
respect to the chemical potentials of the contacts. As will be
shown, electron and photon fluctuations and their correlations are
strongly sensitive to the concrete configuration of the system.

\section{Resonance Fluorescence limit}
\label{prf}
\begin{figure}[t]
\begin{center}
\includegraphics[width=3in,clip]{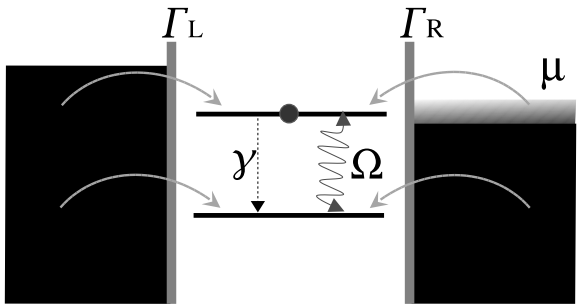}
\end{center}
\caption{\label{esqlb}\small When the chemical potencial in the right lead, $\mu$ is above the energy of both levels, the electron remains in the quantum dot and photons are spontaneously emitted  analogously as photons in resonance fluorescent atoms.
}
\end{figure}
The chemical potential of the left lead is considered to be well above the energies of the QD, so it can be considered infinite. If the chemical potencial of the right lead, $\mu$, is also above the energies of both levels,
 $\mu>\varepsilon_{1(2)}$, the QD is always populated by one electron and transport is cancelled.
 Then, this case is completely analogue to the resonance fluorescence problem,
 where spontaneously emitted photons play the role of fluorescent photons:
 the trapped electron is coherently delocalized by the driving field between
 the two levels performing photo-assisted Rabi oscillations until the emission
 of a photon, then the electron is relaxed to the lower level, cf. Fig. \ref{esqlb}.

We consider a small correction to this behaviour due to the thermal smearing of the Fermi level for finite temperatures. Then, there is a contribution of transport by a small but finite probability for the electron to be extracted to the right lead when it occupies the upper level. The Fermi distribution function weighting this transition can be approximated by $\bar f_2=x\approx e^{\beta(\varepsilon_2-\mu)}$, where $\beta=(k_BT)^{-1}$, see Fig. \ref{esqlb}.
%At finite temperature, the Fermi level is smeared
%In the case shown in Fig. \ref{esqlb}, 
%though $\epsilon_{2}$ is below $\mu$, thermally activated tunneling can ocur, so the electron in the upper level has a finite probability to be extracted to the collector
%where $\epsilon_{2}$ is below but close to $\mu$, the electron in the upper level has a small probability, $x\Gamma_{\rm R}$, to be extracted to the right lead
%due to the thermal broadening of the Fermi level for finite temperatures. Here, we will consider that this effect is small, so $f_2=1-x$ and $\bar f_2=x\approx e^{\beta(\varepsilon_2-\mu)}$, where $\beta=(k_BT)^{-1}$. 
The effect of thermal smearing on electronic transport through a quantum dot has been controlled recently\cite{maclean}. Then, photons deviate from the resonance
fluorescence like statistics because the QD may be empty during
short lapses of time. It would be the case if the resonance fluorescent atom could be eventually ionized. Then, from the Taylor expansion for low $x$, one obtains a finite electronic current:
\begin{equation}
\label{c10nb}
I_e=\frac{\Omega ^2 \Gamma _{\rm L} \Gamma _{\rm R}}{\left(\gamma ^2+2 \Omega ^2\right) \left(\Gamma _{\rm L}+\Gamma _{\rm R}\right)}x+O\left(x^2\right)
\end{equation}
that introduces a small contribution in the photonic emission:
\begin{equation}
\label{c01nb}
I_p=\frac{\gamma  \Omega ^2}{\gamma ^2+2 \Omega ^2}-\frac{\gamma  \Omega ^2 \Gamma _{\rm R} \left(\Omega ^2+3 \gamma  \left(\Gamma _{\rm L}+\Gamma_{\rm R}\right)\right)}{2 \left(\left(\gamma ^2+2 \Omega ^2\right)^2 \left(\Gamma _{\rm L}+\Gamma _{\rm R}\right)\right)}x+O\left(x^2\right)
\end{equation}

\begin{figure}[t]
\begin{center}
\includegraphics[width=\linewidth,clip]{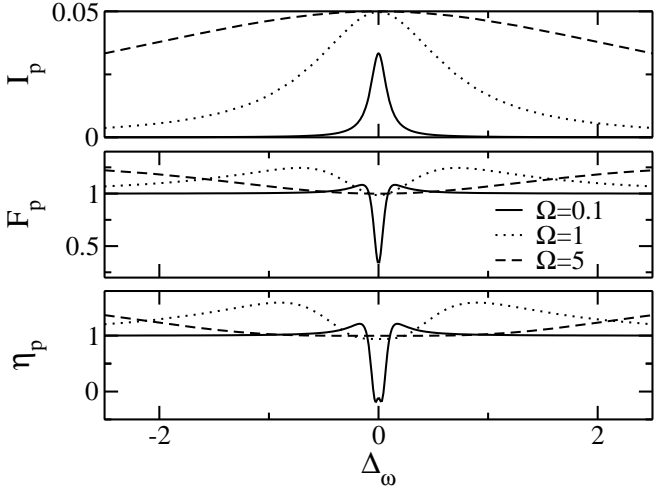}
\end{center}
\caption{\label{phrf}\small {\bf $\mu\gtrsim\varepsilon_2$:} Dependence of the photonic current, Fano factor and skewness with the detuning, for different field intensities in the regime where no levels are in the transport window: $\varepsilon_{1,2}<\mu$. $\Gamma_{\rm L}=\Gamma_{\rm R}=\Gamma=1$, $\gamma=0.1$, $x\approx0$. Since electronic transport is cancelled in this regime, the photonic statistics are equivalent to the resonance fluorescence problem. Sub-Poissonian photonic statistics are only found near resonance. It must be noted here, however, that the validity of these results, obtained within the rotating wave approximation, is guarranteed only for $\Delta_\omega\approx0$.
}
\end{figure}

The photonic resonance fluorescence behaviour as well as
electronic transport quenching is recovered for $x=0$, cf. Fig. \ref{phrf}. There, it can be seen that the sub-Poissonian photon behaviour goes super-Poissonian in the vicinity of the resonance, as described by (\ref{frf}). In those regions, and opposite to what is seen in resonance, the AC intensity increase the deviation of the statistics from the Poissonian values. From (\ref{c10nb}), (\ref{c01nb}) and the
expressions shown in Appendix \ref{aprf} for the second order moments, one obtains the contribution of the thermal smearing of the collector in the electronic statistics (for $\Gamma _{\rm L}=\Gamma _{\rm R}=\Gamma$) and the expected photonic Fano factor:
\begin{eqnarray}
F_e&=&
%1+\Gamma _{\rm L} \Gamma _{\rm R}\left(\frac{\gamma  (\gamma^2 -2 \Omega^2 )}{\left(\gamma ^2+2 \Omega ^2\right)^2 \left(\Gamma _{\rm L}+\Gamma _{\rm R}\right)}-\frac{\Omega ^2}{\left(\Gamma _{\rm L}+\Gamma _{\rm R}\right)^2}\right)x+O\left(x^2\right)\\
1+\frac{1}{4}\left(\frac{2\gamma\Gamma(\gamma^2 -2 \Omega^2 )}{\left(\gamma ^2+2 \Omega ^2\right)^2 }-\Omega ^2\right)x+O\left(x^2\right)\\
F_p&=&1-\frac{6 \gamma ^2\Omega ^2}{\left(\gamma ^2+2 \Omega ^2\right)^2}+O\left(x\right).%\nonumber\\[-2.5mm]
%&&\\[-2.5mm]
%&&-\frac{\gamma\Omega^2\Gamma_{\rm R}}{2\left(\gamma ^2+2 \Omega ^2\right)^2}\left(\frac{\gamma ^3-10 \Omega ^2 \gamma}{\left(\gamma ^2+2 \Omega ^2\right) \left(\Gamma_{\rm L}+\Gamma _{\rm R}\right)}-\frac{4 \left(7 \gamma ^2-4 \Omega ^2\right)}{\gamma ^2+2 \Omega ^2}-\frac{\Omega^2}{\left(\Gamma_{\rm L}+\Gamma _{\rm R}\right)^2}\right)x
%\frac{\left(\gamma  \Omega ^2 \Gamma _{\rm R} \left(\left(\Gamma_{\rm L}+\Gamma _{\rm R}\right) \left(\gamma ^3-10 \Omega ^2 \gamma -4 \left(7 \gamma ^2-4 \Omega ^2\right) \left(\Gamma _{\rm L}+\Gamma_{\rm R}\right)\right)-\Omega ^2 \left(\gamma ^2+2 \Omega ^2\right)\right)\right) x}{2 \left(\left(\gamma ^2+2 \Omega ^2\right)^3 \left(\Gamma_{\rm L}+\Gamma _{\rm R}\right)^2\right)}
%+O\left(x^2\right).\nonumber
\end{eqnarray} 
The driving field induces sub-Poissonian photonic noise which (in
the limit $x=0$) reaches a minimum $F_{p,m}=\frac{1}{4}$ for
$\Omega_{\rm m}=\gamma/\sqrt{2}$ before the Rabi oscillations
dominate the dynamics over relaxation processes, cf. Fig. \ref{nb}.
The electron-photon correlation coefficient becomes (see Appendix
\ref{aprf}):
\begin{eqnarray}
r=
%\left(\gamma\left(\Gamma _{\rm L}+ \Gamma _{\rm R}\right)\left(\gamma ^2-10\Omega ^2\right)-\Omega ^2\left(\gamma ^2+2 \Omega ^2\right)\right)\nonumber\\
%\times\sqrt{\frac{\gamma\Gamma_{\rm L}\Gamma_{\rm R}x}{2\left(\Gamma_{\rm L}+\Gamma _{\rm R}\right)^3\left(\gamma ^4-2 \Omega ^2 \gamma ^2+4 \Omega ^4\right)}}+O\left(x^{3/2}\right),
\left(2\Gamma\gamma\left(\gamma ^2-10\Omega ^2\right)-\Omega ^2\left(\gamma ^2+2 \Omega ^2\right)\right)\nonumber\\
\times\sqrt{\frac{\gamma x}{16\Gamma\left(\gamma ^4-2 \Omega ^2 \gamma ^2+4 \Omega ^4\right)}}+O\left(x^{3/2}\right),
\end{eqnarray}
where it is clear that the AC field contributes to negative electron-photon correlations, cf. Fig. \ref{nb}.
The third cumulants become:
\be
\eta_e{=}
%1+\frac{3 \Gamma _{\rm L} \Gamma _{\rm R} \left(\gamma  (\gamma^2 -4 \Omega^2 ) \left(\Gamma _{\rm L}+\Gamma _{\rm R}\right)-\Omega ^2 \left(\gamma ^2+2\Omega ^2\right)\right) x}{\left(\gamma ^2+2 \Omega ^2\right)^2 \left(\Gamma _{\rm L}+\Gamma _{\rm R}\right)^2}+O\left(x^2\right)\\
%1+\frac{3\left(2\gamma\Gamma  (\gamma^2 -4 \Omega^2)-\Omega ^2 \left(\gamma ^2+2\Omega ^2\right)\right) x}{4\left(\gamma ^2+2 \Omega ^2\right)^2}+O\left(x^2\right)\\
1+\frac{3}{4}\left(\frac{2\gamma\Gamma  (\gamma^2 -4 \Omega^2)}{\left(\gamma ^2+2 \Omega ^2\right)^2}-\frac{\Omega ^2}{\gamma ^2+2 \Omega ^2}\right) x+O\left(x^2\right),
\ee
for electrons, and:
%\\
\be
\eta_p{=}1-\frac{6 \Omega^2\gamma^2\left(3\gamma^4-4 \Omega ^2\gamma ^2+16 \Omega^4\right)}{\left(\gamma ^2+2 \Omega ^2\right)^4}+O\left(x\right),
%+\frac{3 \gamma\Omega ^2 \Gamma _{\rm R} \left(\left(\Gamma _{\rm L}+\Gamma _{\rm R}\right) \left(16 \Omega ^8-20 \gamma ^2 \Omega ^6-8 \gamma ^4 \Omega ^4+3 \gamma ^6\Omega ^2+2 \left(\Gamma _{\rm L}+\Gamma _{\rm R}\right) \left(-\gamma ^7+16 \Omega ^2 \gamma ^5-80 \Omega ^4 \gamma ^3+56 \Omega ^6 \gamma +4 \left(7\gamma ^6-51 \Omega ^2 \gamma ^4+78 \Omega ^4 \gamma ^2-16 \Omega ^6\right) \left(\Gamma _{\rm L}+\Gamma _{\rm R}\right)\right)\right)-\gamma  \Omega ^4\left(\gamma ^2+2 \Omega ^2\right)^2\right) x}{4 \left(\gamma ^2+2 \Omega ^2\right)^5 \left(\Gamma _{\rm L}+\Gamma _{\rm R}\right)^3}+O\left(x^2\right)
\ee
for photons.

Interestingly, the strong photonic noise supression coincides with a region where the skewness almost vanishes, cf. Fig. \ref{nb}a, leading to the possibility to operate the device as a {\it regular boson source}.

Two asymptotic limits of the results presented above, the undriven and high field intensity limits will be considered.
\begin{figure}[t]
\begin{center}
\includegraphics[width=\linewidth,clip]{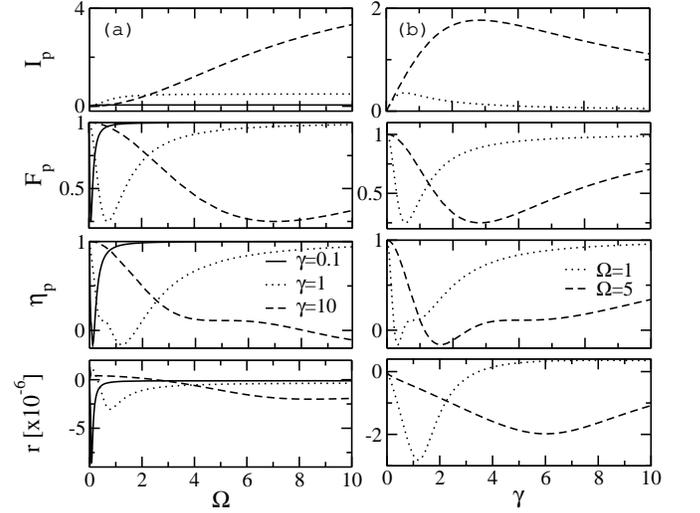}
\end{center}
\caption{\label{nb}\small {\bf $\mu\gtrsim\varepsilon_2$:} Dependence of the photonic current, Fano factor and skewness and the electron-photon correlation with (a) the field
intensity, $\Omega$, for different photon emission rates and (b) the photon
emission rate, $\gamma$ for different field intensities in the regime where no levels are in the transport window: $\varepsilon_{1,2}<\mu$. $\Gamma_{\rm L}=\Gamma_{\rm R}=\Gamma=1$, $x\approx0$. $F_p$ and $\eta_p$ show a pronounced minimum in their dependence with the field intensity typical for resonance fluorescence. In the non driven case, $I_p=0$, $F_p=\eta_p=1$ and $r=0$.
}
\end{figure}

\subsection{Undriven case, $\Omega=0$}
In the absence of driving, once an electron occupies the lower level --by direct tunneling from the leads or by relaxation from the upper one-- there is no process able to remove the electron from the lower level. Then, the stationary state of the system coincides with $\rho_2=1$ and both photon emission and electron tunneling are cancelled:
\begin{equation}
%c_{10}=c_{01}=c_{20}=c_{02}=c_{11}=0.
c_{ij}=0 \quad \forall ~i,j.
\end{equation}
As expected, the cancellation of photon emission makes all the photonic cumulants Poissonian, so $F_p=\eta_p=1$.
However, a small contribution of the tunneling through the upper level modifies the electronic shot noise, so the Fano factor:
\begin{equation}
F_e=1+\frac{2 x \Gamma _{\rm L} \Gamma _{\rm R}}{(\Gamma _{\rm L}+\Gamma _{\rm R}) \left(2 \gamma +x \Gamma _{\rm R}\right)-x\gamma\Gamma _{\rm R}}
\end{equation}
%\stackrel{\Gamma_{\rm L}=\Gamma_{\rm R}}{\longrightarrow}1+\frac{2\Gamma x}{4\gamma+x\Gamma}\\
%F_p&=&1\\
and the skewness of the statistics:
\be
\eta_e=1+\frac{6 x \Gamma _{\rm L} \Gamma _{\rm R} \left(2 \Gamma _{\rm L} \left(\gamma +x \Gamma _{\rm R}\right)+\Gamma _{\rm R} \left(x \Gamma _{\rm R}-(x-2) \gamma\right)\right)}{\left(\Gamma _{\rm L} \left(2 \gamma +x \Gamma _{\rm R}\right)+\Gamma _{\rm R} \left(x \Gamma _{\rm R}-(x-2) \gamma \right)\right)^2}.
%\eta_p&=&1.
\ee
are spuriously super-Poissonian. 
Then, also the electron-photon correlation is affected. For the case $\Gamma_{\rm L}=\Gamma_{\rm R}=\Gamma$:
\be
r=
%&=&\sqrt{\frac{x\gamma\Gamma_{\rm L}\Gamma_{\rm R}\left(2\Gamma_{\rm L}+(2-x)\Gamma_{\rm R}\right)}{2\left(\Gamma_{\rm L}\left(2\gamma+x\Gamma_{\rm R}\right)+\Gamma_{\rm R}\left(x\Gamma_{\rm R}+(2-x)\gamma \right)\right)\left(\Gamma_{\rm R}\left(x\Gamma_{\rm R}+(2-x)\gamma\right)+\Gamma_{\rm L}\left(2\gamma+3x\Gamma_{\rm R}\right)\right)}}\nonumber\\
%&&\stackrel{\Gamma_{\rm L}=\Gamma_{\rm R}}{\longrightarrow}
\sqrt{\frac{(4-x)\gamma x}{2(2x\Gamma+(4-x)\gamma)(4x\Gamma+(4-x)\gamma)}}.
\ee
All these deviations are obviously cancelled as the rate for extracting an electron from the upper level, $x\Gamma_{\rm R}$, goes to zero.

\subsection{High intensity limit: $\Omega\rightarrow\infty$}
Increasing the intensity of the AC field, the electron tends to
occupy the upper level with a probability:
$\rho_{2}=\frac{2\Gamma_{\rm L}+(2-x)\Gamma_{\rm R}}{4\Gamma_{\rm
L}+(4-x)\Gamma_{\rm R}}$ ($\sim\frac{1}{2}$, when $x\rightarrow 0$). Then, it can tunnel to
the right contact (with a probability $x\Gamma_{\rm R}$) or be relaxed to the lower level (with a probability $\gamma$), contributing to finite electronic and photonic currents:
\begin{eqnarray}
I_e&=&\frac{2x\Gamma_{\rm L}\Gamma_{\rm R}}{4\Gamma_{\rm L}+(4-x)\Gamma_{\rm R}}\\
%=&\frac{2x \Omega ^2 \Gamma _{\rm L} \Gamma _{\rm R}}
%{\Gamma _{\rm R} \left((2-x) \gamma ^2+(3-x) x \Gamma _{\rm R}
%\gamma +(4-x) \Omega ^2\right)+\Gamma _{\rm L} \left(2 \gamma ^2+3 x \Gamma _{\rm R} \gamma +
%4 \Omega ^2\right)+x^2\Gamma_{\rm R}^2(\Gamma_{\rm L}+\Gamma_{\rm R})}\nonumber\\
%\stackrel{x\rightarrow0}{\longrightarrow}\frac{x \Omega ^2 \Gamma _{\rm L} \Gamma _{\rm R}}{(\gamma^2+2\Omega^2)(\Gamma_{\rm L}+\Gamma_{\rm R})}
%\end{equation}
%or to a photonic current:
%\begin{equation}
I_p&=&\frac{\gamma(2\Gamma_{\rm L}+(2-x)\Gamma_{\rm R})}{4\Gamma_{\rm L}+(4-x)\Gamma_{\rm R}}.
%\sim\frac{\gamma}{2}.
\end{eqnarray}
The electronic dynamics is then quite similar to the single resonant level case,\cite{hersfield} so the Fano factor becomes slightly sub-Poissonian:
\begin{equation}
F_e=1-\frac{8 x \Gamma _{\rm L} \Gamma _{\rm R}}{\left(4 \Gamma _{\rm L}+(4-x) \Gamma _{\rm R}\right)^2}.
%\stackrel{\Gamma_{\rm L}=\Gamma_{\rm R}}{\longrightarrow}1-\frac{8x}{(8-x)^2}\sim1-\frac{x}{8}.
\end{equation}
% This result should be compared with the single level case, where it is known that $F_e=1-\frac{x}{2}$ and the probability of finding an electron in the QD is $\rho=\frac{2-x}{2}\sim1$, see Appendix \ref{srlfcs}.

Since the occupation probability of the upper level at high field
intensity is maximum, so it is the probability of finding
the QD unoccupied, $\rho_0=\frac{x\Gamma_{\rm R}}{4\Gamma_{\rm
L}+(4-x)\Gamma_{\rm R}}$, after the extraction of the electron to the right lead. This introduces lapses of time when
photon emission is suppressed, affecting the photonic statistics
by turning it super-Poissonian:
\begin{equation}\label{fplbhi}
F_p=1+\frac{2 x \gamma  \Gamma _{\rm R}}{\left(4 \Gamma _{\rm L}+(4-x) \Gamma _{\rm R}\right)^2}.
%\stackrel{\Gamma_{\rm L}=\Gamma_{\rm R}}{\longrightarrow}1+\frac{2\gamma x}{(8-x)^2\Gamma}
\end{equation}
On contrary, electron-photon correlation is negative since the detection
of an electron (photon) reduces the probability of detecting
 a photon (electron): when an electron has tunnelled out of the system (therefore the quantum dot is empty), photon emission is suppressed. On the other hand, when a photon has been emitted, the upper level is unoccupied and no electron can be extracted from the quantum dot. For $\Gamma_{\rm L}=\Gamma_{\rm R}=\Gamma$:
\begin{eqnarray}
%r&=&-\frac{\sqrt{2x \gamma  \Gamma _{\rm L} \Gamma _{\rm R}} }{\sqrt{32 \Gamma _{\rm L}^3+48 (2-x) \Gamma _{\rm R}\Gamma _{\rm L}^2+2 \left(9 x^2-40 x+48\right) \Gamma _{\rm R}^2 \Gamma _{\rm L}+(4-x)^2 (2-x) \Gamma _{\rm R}^3 }}\nonumber\\[-2.5mm]
%&&\\[-2.5mm]
%&&\times\frac{4 \Gamma _{\rm L}+(4-3 x) \Gamma _{\rm R}}{\sqrt{16 \Gamma _{\rm L}^2+8(4-x) \Gamma _{\rm R}\Gamma _{\rm L}+\Gamma _{\rm R} \left(\Gamma _{\rm R} (4-x)^2+2x\gamma \right)}}\nonumber
r=-\frac{\sqrt{2x \gamma}(8-3x) }{\sqrt{(4-x)(64-24x+x^2)(2x\gamma+(8-x)^2\Gamma) }}.
\end{eqnarray}

For the higher moments, one obtains:
\begin{eqnarray}
\eta_e&=&
%1-\frac{24 x \Gamma _{\rm L} \Gamma _{\rm R} \left(16 \Gamma _{\rm L}^2+16 (2-x) \Gamma _{\rm R} \Gamma _{\rm L}+(4-x)^2 \Gamma _{\rm R}^2\right)}{\left(4 \Gamma _{\rm L}+(4-x) \Gamma _{\rm R}\right)^4}\\
1-\frac{24x\left(64-\left(24-x\right)x\right)}{\left(8-x\right)^4}\\
\eta_p&=&
%1+\frac{6 x \gamma  \Gamma _{\rm R} \left(16 \Gamma _{\rm L}^2+4\left(2 (4-x) \Gamma _{\rm R}-\gamma\right) \Gamma _{\rm L}+\Gamma _{\rm R} \left(\Gamma _{\rm R} (4-x)^2-(4-3x)\gamma \right)\right)}{\left(4 \Gamma _{\rm L}+(4-x) \Gamma _{\rm R}\right)^4}
1+\frac{6x\gamma\left((8-x)^2\Gamma-(8-3x)\gamma\right)}{(8-x)^4 \Gamma^2}.
\end{eqnarray}

%See Figs. \ref{nb}, \ref{mug01}, \ref{mug1} and \ref{gmu}.*****

\section{Dynamical Channel Blockade regime}
\label{secdcb}

\begin{figure}[t]
\begin{center}
\includegraphics[width=3in,clip]{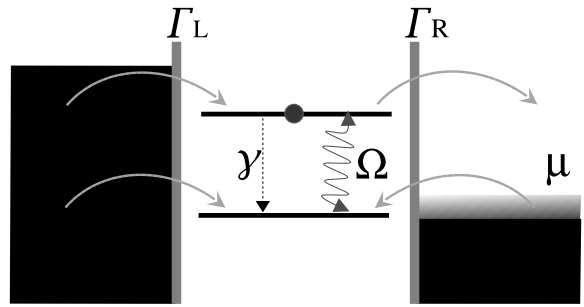}
\end{center}
\caption{\label{esqdcb}\small Dynamical channel blockade
configuration, where the electronic transport is strongly
suppressed through the lower level, though there is a small probability introduced by thermal smearing of the Fermi surface in the right lead. Again, the chemical potencial of the left lead is considered infinite.}
\end{figure}
If the chemical potential of the right lead lies between the energy levels of the QD,
 $\varepsilon_1<\mu<\varepsilon_2$ and therefore, $f_1=1-x$, $f_2=0$,
 where $x\approx e^{\beta(\varepsilon_1-\mu)}$ and $\beta=(k_BT)^{-1}$, electronic transport
 is strongly suppressed through the lower level, cf. Fig. \ref{esqdcb}.
 Then, since only one electron is allowed in the system, the occupation of the lower level avoids the entrance of electrons from the left lead and the current is blocked.
 This mechanism, which is known as {\it dynamical channel blockade}, predicts electronic super-Poissonian
  shot noise in multichannel systems like, for instance, two-level quantum dots\cite{belzig} or capacitively coupled
  double quantum dots\cite{gattobigio,yoHaug} as well as positive cross-correlations in three terminal
  devices\cite{cottet3}. It has been proposed as the responsible of noise enhancement
measured experimentally in multilevel quantum dots\cite{zhang} and stacks of double quantum dots\cite{barthold}.

\begin{figure}[t]
\begin{center}
\includegraphics[width=\linewidth,clip]{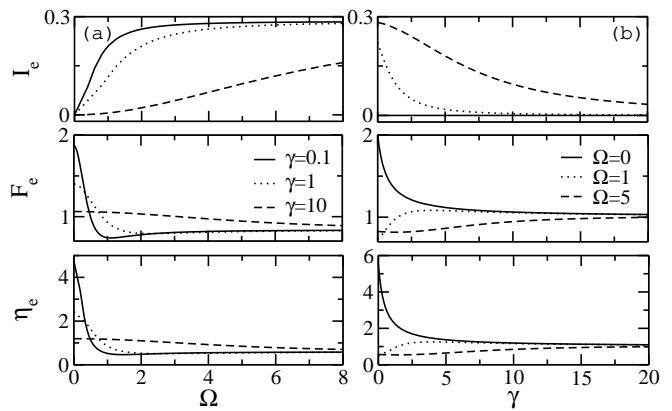}
\end{center}
\caption{\label{edcb}\small {\bf Dynamical Channel Blockade:}
Dependence of $I_e$, $F_e$ and $\eta_e$ with
(a) the field intensity, $\Omega$, for different photon emission rates,
(b) the photon emission rate, $\gamma$ for different field intensities, in the dynamical channel blockade regime:
$\varepsilon_{1}<\mu<\varepsilon_{2}$. $\mu<\varepsilon_{1,2}$.
$\Gamma_{\rm L}=\Gamma_{\rm R}=\Gamma=1$,
$\tilde\Omega=\Omega/\Gamma$, $\tilde\gamma=\gamma/\Gamma$,
$x\approx0$. As discussed in the text, the super-Poissonian electronic Fano factor typical for dynamical channel blockade, is turned sub-Poissonian by the AC intensity. }
\end{figure}
\begin{figure}[t]
\begin{center}
\includegraphics[width=\linewidth,clip]{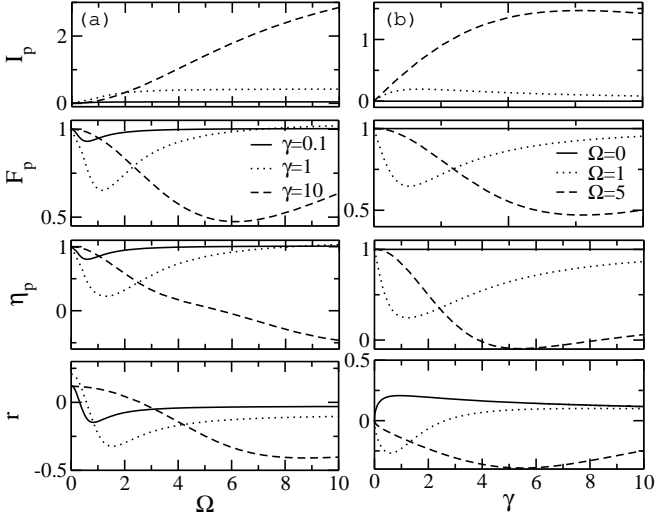}
\end{center}
\caption{\label{phdcb}\small {\bf Dynamical Channel Blockade:}
Dependence of the photonic current, Fano factor and skewness and the electron-photon correlation coefficient with (a) the field intensity, for different photon emission rates and (b) the photon emission rate, $\gamma$, for different field intensities in the dynamical channel blockade regime:
$\varepsilon_{1}<\mu<\varepsilon_{2}$. $\mu<\varepsilon_{1,2}$.
$\Gamma_{\rm L}=\Gamma_{\rm R}=\Gamma=1$, $x\approx0$.}
\end{figure}
The blocking of the current is not forever since the electron in
the lower level has a finite but small probability of tunneling to
the collector, $x\Gamma_{\rm R}$, due to the thermal smearing of
the Fermi level. Then, the trapped electron eventually escapes to
the right lead allowing electrons to tunnel through the upper
level before the lower one is again occupied. Thus, the current is
restricted to short lapses of time while for long periods $t\sim
(x\Gamma_{\rm R})^{-1}$ transport is quenched. This {\it bunching}
of electrons is reflected in super-Poissonian shot noise.

Photon-mediated relaxation introduces an additional way to occupy the lower level when current is flowing through the upper one, shortening the lapse of time when transport is allowed. Thus, the electrons are transferred in smaller bunches and the super-Poissonian character of the electronic noise is reduced. The detection of a photon is always at the end of a bunch of electrons and implies the cancelation of the current, leading to a positive electron-photon correlation.

The introduction of the AC field pumps the electron in the lower state to the upper one, giving the electron a finite probability to tunnel to the right lead or to be relaxed by the emission of one photon.
This reduces the electronic shot noise by reducing the duration of the lapses of time when transport is blocked (opposite to the effect of photons). Thus, when $x=0$, the electronic current and the photonic emission are proportional to the driving intensity:
\begin{eqnarray}
I_e=\frac{2 \Omega ^2 \Gamma _{\rm L} \Gamma _{\rm R}}{\Gamma
_{\rm R} \left(\tilde\Gamma _{\rm R}^2 +3 \Omega
^2\right)+\Gamma _{\rm L} \left(\tilde\Gamma _{\rm R}(2\gamma+\Gamma _{\rm R}) +4 \Omega ^2\right)}
%\stackrel{\Gamma_{\rm L}=\Gamma_{\rm R}}{\longrightarrow}
%\frac{2 \Gamma  \Omega ^2}{7 \Omega ^2+(\gamma +\Gamma ) (3 \gamma +2 \Gamma )}
\\
I_p=\frac{\gamma  \Omega ^2 \left(2 \Gamma _{\rm L}+\Gamma
_{\rm R}\right)}{\Gamma
_{\rm R} \left(\tilde\Gamma _{\rm R}^2 +3 \Omega
^2\right)+\Gamma _{\rm L} \left(\tilde\Gamma _{\rm R}(2\gamma+\Gamma _{\rm R}) +4 \Omega ^2\right)}
%\stackrel{\Gamma_{\rm L}=\Gamma_{\rm R}}{\longrightarrow}
%\frac{3 \gamma  \Omega ^2}{7 \Omega ^2+(\gamma +\Gamma ) (3 \gamma +2 \Gamma )}
\end{eqnarray}
and channel blockade is removed, see Figs. \ref{edcb} and \ref{phdcb}. We have defined $\tilde\Gamma_{\rm R}=\gamma+\Gamma_{\rm R}$. Considering, for simplicity, the case $\Gamma_{\rm L}=\Gamma_{\rm R}=\Gamma$ and $\tilde\Gamma=\gamma+\Gamma$, the Fano factors become (see Appendix \ref{apdcb}):
%\begin{eqnarray}
%F_e&=&1-\frac{8 \Omega ^4+2 \left(2 \gamma ^2+15 \Gamma  \gamma +9 \Gamma ^2\right) \Omega ^2-2 \Gamma  (\gamma +\Gamma )^2 (3 \gamma +2 \Gamma)}{\left(7 \Omega ^2+(\gamma +\Gamma ) (3 \gamma +2 \Gamma )\right)^2}\\
%F_p&=&1-\frac{2 \gamma  \Omega ^2 \left(22 \Gamma ^2+28 \gamma  \Gamma -\Omega ^2\right)}{\Gamma  \left(7 \Omega ^2+(\gamma +\Gamma ) (3 \gamma +2\Gamma )\right)^2}.
%\end{eqnarray}
\be
F_e=1-\frac{8 \Omega ^4+2 \left(2 \gamma ^2+15 \Gamma  \gamma +9 \Gamma ^2\right) \Omega ^2-2 \Gamma\tilde\Gamma^2 (3 \gamma +2 \Gamma)}{\left(7 \Omega ^2+\tilde\Gamma(3 \gamma +2 \Gamma )\right)^2},
\ee
for electrons, and:
\be
F_p=1-\frac{2 \gamma  \Omega ^2 \left(22 \Gamma ^2+28 \gamma  \Gamma -\Omega ^2\right)}{\Gamma  \left(7 \Omega ^2+\tilde\Gamma(3 \gamma +2\Gamma )\right)^2},
\ee
for photons. As can be seen in Fig. \ref{phdcb}a, the minimun that appeared in the resonance fluorescence configuration still appears, but its depth and position now depend on the tunneling rates. The modification of the resonance fluorescence behaviour is also reflected in the super-Poissonian large AC intensities asymptotic value (discussed below).

The electron-photon correlation coefficient will be considered in the asymptotic non-driven and high field intensity cases. As expected, the driving contributes to make the electronic noise sub-Poissonian and the photonic one super-Poissonian. However, it has to compete with the photon emission that contributes to bring the electron to the lower state and to block the current. In Fig. \ref{edcb}b, it can be seen how the pumped electronic current is decreased by the photon emission rate and the Fano factor tends asymptotically to be Poissonian. The positive electron-photon correlation is decreased by the ac field since the emission of a photon does not imply transport blocking anymore, as seen in Fig. \ref{phdcb}.

\subsection{Undriven case, $\Omega=0$}
The most interesting
features appear in the absence of the ac field, where the
consequences of the dynamical channel blockade are maximal and
there is a strong dependence of the statistics on the thermal
smearing factor, $x$. In the absence of photons ($\gamma=0$), the electronic
current and Fano factor are:
\begin{eqnarray}
I_e&=&\frac{2x\Gamma_{\rm L}\Gamma_{\rm R}}{(x+1)\Gamma_{\rm L}+\Gamma_{\rm R}}\\
%\approx\frac{2x\Gamma_{\rm L}\Gamma_{\rm R}}{\Gamma_{\rm L}+\Gamma_{\rm R}}\\
F_e&=&1+\frac{2\Gamma_{\rm L}\left((1-x)^2\Gamma_{\rm L}+(1-3x)\Gamma_{\rm R}\right)}{((x+1)\Gamma_{\rm L}+\Gamma_{\rm R})^2}.
%\approx 1+\frac{2\Gamma_{\rm L}}{\Gamma_{\rm L}+\Gamma_{\rm R}}.
\end{eqnarray}
It is interesting to see here how the Fano factor can be tuned by
the asymmetric coupling to the leads: $F_e=3$ (if $\Gamma_{\rm
L}\gg\Gamma_{\rm R}$), $F_e=2$ (if $\Gamma_{\rm L}=\Gamma_{\rm
R}$) and $F_e=1$ (if $\Gamma_{\rm L}\ll\Gamma_{\rm R}$). In the
lastest case, the contribution of $x\Gamma_{\rm R}$ is diminished and
the left barrier controls the transport (in this limit, the
current is $I_e=2x\Gamma_{\rm L}$). Then, the transferred
electrons are uncorrelated one from the others resembling the
behaviour of the single barrier problem briefly discussed above. The case $\Gamma_{\rm L}\gg\Gamma_{\rm R}$ was studied in Ref. \cite{belzig} without considering the processes that introduce an electron from the collector to the lower level, with a rate $(1-x)\Gamma_{\rm R}$. These transitions do not contribute in this particular limit, but that is not the case for the rest of configurations.

Considering photon emission and small $x$, one can expand the
first coefficients for the electronic and photonic statistics, as
well as for the electron-photon correlations. Relaxation by photons contribute to shorten the bunches of electrons flowing through the upper level when the lower one is empty, thus reducing both the electronic current:
\be
I_e=\frac{2 \Gamma _{\rm L} \Gamma _{\rm R} \left(\gamma +\Gamma _{\rm R}\right) x}{\Gamma _{\rm R} \left(\gamma +\Gamma _{\rm R}\right)+\Gamma _{\rm L} \left(2 \gamma +\Gamma_{\rm R}\right)}+O\left(x^2\right)
%\stackrel{\Gamma_{\rm L}=\Gamma_{\rm R}}{\longrightarrow}2x\Gamma\frac{\gamma+\Gamma}{3\gamma+2\Gamma}+O\left(x^2\right)
\ee
and Fano factor ({\it noise reduction by noise}):
\bea
F_e&{=}&1+\frac{2 \Gamma _{\rm L} \Gamma _{\rm R}}{\Gamma _{\rm R} \tilde\Gamma _{\rm R}+\Gamma _{\rm L} \left(\gamma +\tilde\Gamma _{\rm R}\right)}\nonumber\\[-2.5mm]
%-\frac{2\Gamma _{\rm L}\Gamma _{\rm R} \left(2 \gamma ^2+5 \Gamma _{\rm R} \gamma +3 \Gamma _{\rm R}^2+2 \Gamma _{\rm L} \left(\gamma +2 \Gamma _{\rm R}\right)\right) x}{\left(\Gamma _{\rm R}\left(\gamma +\Gamma _{\rm R}\right)+\Gamma _{\rm L} \left(2 \gamma +\Gamma _{\rm R}\right)\right)^2}+O\left(x^2\right)
\\[-2.5mm]
&&-2\Gamma _{\rm L}\Gamma _{\rm R}\frac{ \tilde\Gamma _{\rm R}^2+\left(\tilde\Gamma _{\rm R}+2\Gamma _{\rm L}\right) \left(\gamma +2 \Gamma _{\rm R}\right) }{\left(\Gamma _{\rm R}\tilde\Gamma _{\rm R}+\Gamma _{\rm L} \left(\gamma +\tilde\Gamma _{\rm R}\right)\right)^2}x+O\left(x^2\right)\nonumber
%&&\stackrel{\Gamma_{\rm L}=\Gamma_{\rm R}}{\longrightarrow}1+\frac{2\Gamma}{3\gamma+2\Gamma}+O(x)
\eea
without affecting to its super-Poissonian character, cf. Fig. \ref{edcb}b. Again, we defined, for simplicity, $\tilde\Gamma _{\rm R}=\gamma+\Gamma _{\rm R}$. The photonic current and Fano factor become:
\begin{eqnarray}
I_p&=&\frac{\gamma  \Gamma _{\rm L} \Gamma _{\rm R} x}{\Gamma _{\rm R} \left(\gamma +\Gamma _{\rm R}\right)+\Gamma _{\rm L} \left(2 \gamma +\Gamma _{\rm R}\right)}+O\left(x^2\right)\\
%\stackrel{\Gamma_{\rm L}=\Gamma_{\rm R}}{\longrightarrow}\frac{\gamma x\Gamma}{3\gamma+2\Gamma}\\
F_p&=&1-\frac{2 \gamma  \Gamma _{\rm L} \Gamma _{\rm R} \left(\gamma +2 \Gamma _{\rm L}+2 \Gamma _{\rm R}\right) x}{\left(\Gamma _{\rm R} \left(\gamma +\Gamma_{\rm R}\right)+\Gamma _{\rm L} \left(2 \gamma +\Gamma _{\rm R}\right)\right)^2}+O\left(x^2\right),
\end{eqnarray}
%\stackrel{\Gamma_{\rm L}=\Gamma_{\rm R}}{\longrightarrow}1-\frac{2\gamma(\gamma+4\Gamma)x}{(3\gamma+2\Gamma)^2}+O(x^2)\\
while we obtain, for the electron-photon correlation coefficient (in the case $\Gamma_{\rm L}=\Gamma_{\rm R}=\Gamma$):
\be
r=
%(2\Gamma_{\rm L}+\Gamma_{\rm R})\sqrt{\frac{\gamma(\gamma +\Gamma_{\rm R})}{2\left(\Gamma _{\rm R} \left(\gamma+\Gamma _{\rm R}\right)+\Gamma _{\rm L} \left(2 \gamma +\Gamma _{\rm R}\right)\right)\left(\Gamma _{\rm R} \left(\gamma +\Gamma _{\rm R}\right)+\Gamma _{\rm L} \left(2 \gamma +3 \Gamma _{\rm R}\right)\right)}}+O\left(x\right)\nonumber\\&&\stackrel{\Gamma_{\rm L}=\Gamma_{\rm R}}{\longrightarrow}
3\sqrt{\frac{\gamma(\gamma+\Gamma)}{(3\gamma+2\Gamma)(3\gamma+4\Gamma)}}+O(x).
\ee
Interestingly, the electron-photon correlation is roughly independent from $x$, which allows to extract information that is not provided by the {\it flat} photonic Fano factor, cf. Fig. \ref{phdcb}b. 

The expected %super-Poissonian electronic statistics and 
positive electron-photon correlation are obtained.
% Interestingly, the presence of photons reduces the  electronic Fano factor ({\it noise reduction by noise}), but does not affect to its super-Poissonian character.
On the other hand, the presence of electronic transport affects the Poissonian photonic statistics by introducing a sub-Poissonian component: once a photon has been emitted, the electron is relaxed to the lower level blocking the transport. A second photon will not be detected until the electron tunnels to the collector and another one enters the upper level, so photonic events are well separated in time.

From the third cumulants, one obtains:
\begin{eqnarray}
\eta_e&=&1+\frac{6 \Gamma _{\rm L} \Gamma _{\rm R} \left(\gamma +\Gamma _{\rm R}\right) \left(2 \Gamma _{\rm L}+\Gamma _{\rm R}\right)}{\left(\Gamma _{\rm R} \left(\gamma +\Gamma_{\rm R}\right)+\Gamma _{\rm L} \left(2 \gamma +\Gamma _{\rm R}\right)\right)^2}+O\left(x\right)\\
%&&-\frac{6 \left(\Gamma _{\rm L} \Gamma _{\rm R} \left(2 \left(2 \gamma ^2+9 \Gamma _{\rm R}\gamma +8 \Gamma _{\rm R}^2\right) \Gamma _{\rm L}^2+\left(4 \gamma ^3+20 \Gamma _{\rm R} \gamma ^2+33 \Gamma _{\rm R}^2 \gamma +17 \Gamma _{\rm R}^3\right) \Gamma_{\rm L}+\Gamma _{\rm R} \left(\gamma +\Gamma _{\rm R}\right)^2 \left(2 \gamma +3 \Gamma _{\rm R}\right)\right)\right) x}{\left(\Gamma _{\rm R} \left(\gamma +\Gamma_{\rm R}\right)+\Gamma _{\rm L} \left(2 \gamma +\Gamma _{\rm R}\right)\right)^3}+O\left(x^2\right)\\
\eta_p&=&1-\frac{6 \left(\gamma  \Gamma _{\rm L} \Gamma _{\rm R} \left(\gamma +2 \Gamma _{\rm L}+2 \Gamma _{\rm R}\right)\right) x}{\left(\Gamma _{\rm R} \left(\gamma +\Gamma_{\rm R}\right)+\Gamma _{\rm L} \left(2 \gamma +\Gamma _{\rm R}\right)\right)^2}+O\left(x^2\right).
\end{eqnarray}

\subsection{High intensity limit: $\Omega\rightarrow\infty$}

If the intensity of the driving field is large enough, the dynamical channel blockade is completely lifted, finding electronic and photonic currents:
\be
\frac{I_e}{2(1+x)\Gamma_{\rm L} \Gamma_{\rm R}}=\frac{I_p}{\gamma\left(2\Gamma_{\rm L}+(1-x)\Gamma_{\rm R}\right)}=\frac{1}{4\Gamma_{\rm L}+(3-x)\Gamma_{\rm R}}
\ee
 so sub-Poissonian electronic noise and super-Poissonian photonic noise are recovered:
\begin{eqnarray}
F_e&=&1-\frac{8\Gamma_{\rm L} \Gamma_{\rm R}}{\left(4 \Gamma _{\rm L}+3 \Gamma _{\rm R}\right)^2}+O(x)\\
%-\frac{8 \left(\Gamma _{\rm L} \Gamma _{\rm R} \left(4 \Gamma _{\rm L}+5 \Gamma_{\rm R}\right)\right) x}{\left(4 \Gamma _{\rm L}+3 \Gamma _{\rm R}\right)^3}+O\left(x^2\right)
%\stackrel{\Gamma_{\rm L}=\Gamma_{\rm R}}{\longrightarrow}\frac{41}{49}+O(x)\\
\label{fpdcbhi}
F_p&=&1+\frac{2 \gamma  \Gamma _{\rm R}}{\left(4 \Gamma _{\rm L}+3 \Gamma _{\rm R}\right)^2}+O(x).
%+\frac{2 \gamma  \Gamma _{\rm R} \left(4 \Gamma _{\rm L}+5 \Gamma _{\rm R}\right) x}{\left(4\Gamma _{\rm L}+3 \Gamma _{\rm R}\right)^3}+O\left(x^2\right)
%\stackrel{\Gamma_{\rm L}=\Gamma_{\rm R}}{\longrightarrow}1+\frac{2\gamma}{49\Gamma}+O(x).
\end{eqnarray}
Interestingly, in this regime, the photonic influence is washed out from the electronic statistics.
Also, the AC field allows the extraction through the upper level of an electron that has been relaxed by the emission of one photon. This means that the electron-photon correlation becomes negative (for $\Gamma_{\rm L}=\Gamma_{\rm R}=\Gamma$):
\begin{eqnarray}
r=
%&=&-\frac{\sqrt{2\gamma  \Gamma _{\rm L} \Gamma _{\rm R}} \left(4 \Gamma _{\rm L}+\Gamma _{\rm R}\right)}{\sqrt{\left(32 \Gamma _{\rm L}^3+48 \Gamma _{\rm R} \Gamma _{\rm L}^2+34\Gamma _{\rm R}^2 \Gamma _{\rm L}+9 \Gamma _{\rm R}^3\right) \left(16 \Gamma _{\rm L}^2+24 \Gamma _{\rm R} \Gamma _{\rm L}+\Gamma _{\rm R} \left(2 \gamma +9 \Gamma _{\rm R}\right)\right)}}+O\left(x\right)\nonumber\\
%&&\stackrel{\Gamma_{\rm L}=\Gamma_{\rm R}}{\longrightarrow}
-5\sqrt{\frac{2\gamma}{123(2\gamma+49\Gamma)}}+O(x).
\end{eqnarray}

Then, by tuning the driving intensity, one can manipulate the character of the shot noise of electrons and photons, turning the super(sub)-Poissonian statistics to sub(super)-Poissonian for electrons(photons) when increasing $\Omega$.

Higher moments are also obtained,
%\begin{eqnarray}
%c_{30}&=&\frac{64 (x+1)^3 \Gamma _{\rm L}^3 \Gamma _{\rm R}^3}{\left(4 \Gamma _{\rm L}+(3-x) \Gamma _{\rm R}\right)^5}\\
%c_{03}&=&-\frac{(x+1) \gamma ^3 \Gamma _{\rm R} \left(8 \Gamma_{\rm L}^2+2 (3-5 x) \Gamma _{\rm R} \Gamma _{\rm L}+\left(3 x^2-4x+1\right) \Gamma _{\rm R}^2\right)}{\left(4 \Gamma_{\rm L}-(x-3)\Gamma _{\rm R}\right)^5}.
%\end{eqnarray}
giving:
\begin{eqnarray}
\eta_e&=&
%1-\frac{24 (x+1) \Gamma _{\rm L} \Gamma _{\rm R} \left(16 \Gamma _{\rm L}^2-16 (x-1) \Gamma _{\rm R} \Gamma _{\rm L}+(x-3)^2 \Gamma _{\rm R}^2\right)}{\left((x-3) \Gamma _{\rm R}-4\Gamma _{\rm L}\right)^4}\\
1-\frac{24 (x+1)\left(41-(22-x)x\right)}{(7-x)^4}\\
\eta_p&=&
%1+\frac{6 (x+1) \gamma  \Gamma _{\rm R} \left(16 \Gamma _{\rm L}^2-4 \left(\gamma +2 (x-3) \Gamma _{\rm R}\right) \Gamma _{\rm L}+\Gamma _{\rm R} \left(\Gamma _{\rm R} (x-3)^2+(3x-1) \gamma \right)\right)}{\left((x-3) \Gamma _{\rm R}-4 \Gamma _{\rm L}\right)^4}.
1+\frac{6 (x+1) \gamma \left((7-x)^2\Gamma-(5-3x)\gamma\right)}{(7-x)^4\Gamma^2}.
\end{eqnarray}

\section{Both levels in the transport window regime}
\label{sechb}
%: $f_1=f_2=0$.}

\begin{figure}[t]
\begin{center}
\includegraphics[width=3in,clip]{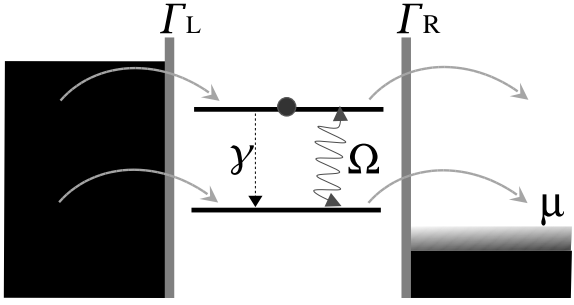}
\end{center}
\caption{\label{esqhb}\small System configuration discussed in section \ref{sechb}, with the two levels in the transport window, $\varepsilon_1,\varepsilon_2>\mu$.
}
\end{figure}
If the energy of both levels are above $\mu$, $\varepsilon_1,\varepsilon_2>\mu$ ($f_1=f_2=0$), the two of them contribute to electronic transport, cf. Fig. \ref{esqhb}. In this particular case, quantum interference effects may be important\cite{gurvitzTL} depending on the concrete geometry of the system. However, in the weak coupling and high frequency limit case considered here, $\varepsilon_2-\varepsilon_1\gg\Gamma_{\rm L,R}$, they can be disregarded. %Then, our system is completely similar to a two resonant levels system.

Contrary to the previous regimes, the contribution of the empty state:
\begin{equation}
\rho_0=\frac{\Gamma_{\rm R}}{2\Gamma_{\rm L}+\Gamma_{\rm R}}
\end{equation}
plays an important role here. It strongly affects the sub-Poissonian character of the photonic noise.

\begin{figure}[t]
\begin{center}
\includegraphics[width=\linewidth,clip]{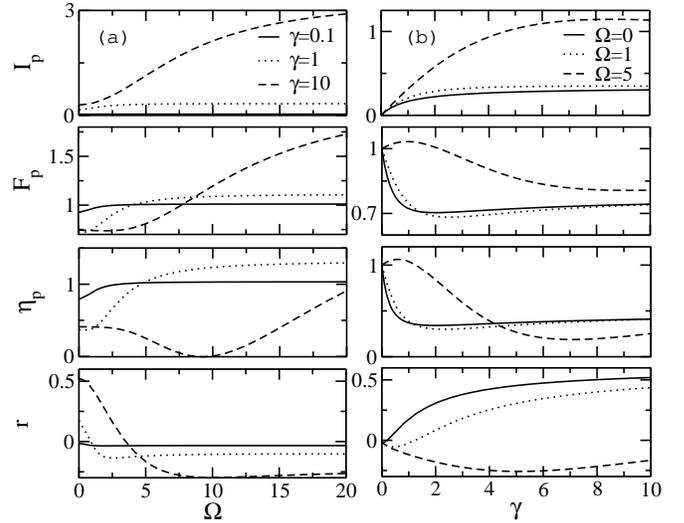}
\end{center}
\caption{\label{hb}\small {\bf $\varepsilon_1,\varepsilon_2>\mu$:} Dependence of the photonic current, Fano factor and skewness and the electron-photon correlation coefficient on (a) the field intensity, $\Omega$, for different photon emission rates, and (b) the photon emission rate, $\gamma$ for different field intensities for $\mu<\varepsilon_{1,2}$. $\Gamma_{\rm L}=\Gamma_{\rm R}=\Gamma=1$. The electronic statistics (not shown) is sub-Poissonian and not affected by the AC field nor by photonic relaxation. The electron-photon correlation coefficient is positive if $\gamma>\Gamma$.
}
\end{figure}

Since the tunneling rates are considered independent on the
energy, electronic transport does not depend on the level that the
electron occupies when tunneling through the QD. Then, the
transport characteristics (electronic current and noise) are
independent of the field intensity, detunning and the spontaneous
emission:
\begin{eqnarray}
\label{iehb}
I_e&=&\frac{2\Gamma_{\rm L}\Gamma_{\rm R}}{2\Gamma_{\rm L}+\Gamma_{\rm R}}\\
%S_e&=&\frac{2\Gamma_{\rm L}\Gamma_{\rm R}(4\Gamma_{\rm L}^2+\Gamma_{\rm R}^2)}{(2\Gamma_{\rm L}+\Gamma_{\rm R})^3}\\
\label{fehb}
F_e&=&\frac{4\Gamma_{\rm L}^2+\Gamma_{\rm R}^2}{(2\Gamma_{\rm L}+\Gamma_{\rm R})^2}.
%\stackrel{\Gamma_{\rm L}=\Gamma_{\rm R}}{\longrightarrow}\frac{5}{9}.
\end{eqnarray}
This case is similar to the single resonant level with a factor 2 in the tunneling from the collector, reflecting that an electron in the left lead finds two different possibilities before tunneling into the QD. 
Similarly to the single resonant level, the Fano factor
is sub-Poissonian. However, the contribution of the two levels
increases the noise.
The normalized third cumulant becomes (see Appendix \ref{aphb}):
\begin{equation}
\eta_e=1-\frac{12 \Gamma _{\rm L} \Gamma _{\rm R} \left(4 \Gamma
_{\rm L}^2+\Gamma _{\rm R}^2\right)}{\left(2 \Gamma _{\rm
L}+\Gamma _{\rm R}\right)^4}.
%\stackrel{\Gamma_{\rm L}=\Gamma_{\rm R}}{\longrightarrow}\frac{7}{27}.
\end{equation}
Interestingly, the two resonant levels statistics coincides with the single resonant one when
 writing $\Gamma_{\rm L}/2$ for $\Gamma_{\rm L}$.
That is not the case for the photonic statistics, that depends on the population of the upper level and,
therefore, on the AC field parameters. For instance, the photonic curret is:
\begin{equation}
I_p=\frac{\gamma\Gamma_{\rm L}\left(2\Omega^2+\Gamma_{\rm
R}(\gamma+2\Gamma_{\rm R})\right)}{(2\Gamma_{\rm L}+\Gamma_{\rm
R})\left(\gamma^2+2\Omega^2+3\gamma\Gamma_{\rm R}+2\Gamma_{\rm
R}^2\right)}.
\end{equation}
The expressions for the second order moments are quite lengthy, unless one considers a simpler case, where the tunneling rates are the same through both barriers, $\Gamma_{\rm L}=\Gamma_{\rm R}=\Gamma$. Then, one obtains a sub-Poissonian Fano factor:
%\begin{equation}
%S_p&=&\frac{\gamma  \left(4 (\gamma +9 \Gamma ) \Omega ^4+2 \left(\gamma ^3+\Gamma  \gamma ^2+40 \Gamma ^2 \gamma +36 \Gamma ^3\right) \Omega^2+\Gamma  (\gamma +2 \Gamma )^2 \left(7 \gamma ^2+10 \Gamma  \gamma +9 \Gamma ^2\right)\right) \left(\gamma  \Gamma +2 \left(\Gamma^2+\Omega ^2\right)\right)}{27 \Gamma  \left(2 \Omega ^2+(\gamma +\Gamma ) (\gamma +2 \Gamma )\right)^3}\\
\bea
%F_p=1-2 \gamma\frac{\Gamma  (\gamma +2 \Gamma )^2 (\gamma +4\Gamma )+\left(14 \Gamma ^2+17\Gamma\gamma-\gamma^2\right)\Omega ^2-2 \Omega ^4}{9 \Gamma  \left(2 \Omega ^2+(\gamma +\Gamma ) (\gamma +2 \Gamma )\right)^2}.
F_p&=&1-\frac{2 \gamma}{9 \Gamma  \left(2 \Omega ^2+(\gamma +\Gamma ) (\gamma +2 \Gamma )\right)^2}\nonumber\\[-1.5mm]
&&\\[-1.5mm]
&&\!\!\!\!\!\!\!\!\!\!\!\!\!\!\!\!\!\!\!\!\times\left(\Gamma  (\gamma +2 \Gamma )^2 (\gamma +4\Gamma )+\left(14 \Gamma ^2+17\Gamma\gamma-\gamma^2\right)\Omega ^2-2 \Omega ^4\right).\nonumber
\eea
%\end{equation}
which can be tuned to super-Poissonian for high enough
intensities. The electron-photon correlation, obtained from (\ref{hbeph}),
%\be
%r=\frac{\sqrt{\gamma } \left(-4 \Omega ^4-2 \left(\gamma ^2+16 \Gamma  \gamma +4 \Gamma ^2\right) \Omega ^2+(5 \gamma -\Gamma ) \Gamma  (\gamma+2 \Gamma )^2\right)}{\sqrt{10} \sqrt{2 \Omega ^2+(\gamma +\Gamma ) (\gamma +2 \Gamma )} \sqrt{\left(4 (\gamma +9 \Gamma ) \Omega ^4+2\left(\gamma ^3+\Gamma  \gamma ^2+40 \Gamma ^2 \gamma +36 \Gamma ^3\right) \Omega ^2+\Gamma  (\gamma +2 \Gamma )^2 \left(7 \gamma ^2+10\Gamma  \gamma +9 \Gamma ^2\right)\right) \left(\gamma  \Gamma +2 \left(\Gamma ^2+\Omega ^2\right)\right)}}
%\ee
may be positive or negative depending on the concrete parametrization of the system, as discussed below. In concrete, positive correlation is obtained when $\Gamma_{\rm L}\ll\Gamma_{\rm R}$ as well as, for low intensity driving,  when the tunneling rates are small compared to the photon emission rate, cf. Fig. \ref{hb}.

\subsection{Undriven case, $\Omega=0$}
The emission of a photon, in this case, depends on the tunneling
of an electron from the left lead to the upper level. Then, it can
tunnel to the collector directly or after being relaxed to the
lower level by the emission of one photon. Therefore, photons {\it
adopt} the electronic sub-Poissonian statistics:
%. The first moments (see Appendix \ref{aphbuc}) give a photonic current and Fano factor:
\begin{eqnarray}
%I_p&=&\frac{\gamma  \Gamma _{\rm L} \Gamma _{\rm R}}{\left(\gamma +\Gamma _{\rm R}\right) \left(2 \Gamma _{\rm L}+\Gamma _{\rm R}\right)} \\
%\stackrel{\Gamma_{\rm L}=\Gamma_{\rm R}}{\longrightarrow}\frac{\gamma\Gamma}{3(\gamma+\Gamma)}\\
%S_p&=&\frac{\gamma  \Gamma _{\rm L} \Gamma _{\rm R} \left(4 \left(\gamma ^2+\Gamma _{\rm R} \left(\gamma +\Gamma _{\rm R}\right)\right)\Gamma_{\rm L}^2+2 \Gamma _{\rm R} \left(\gamma^2+2 \Gamma _{\rm R} \left(\gamma +\Gamma _{\rm R}\right)\right) \Gamma _{\rm L}+\Gamma _{\rm R}^2 \left(\gamma +\Gamma _{\rm R}\right)^2\right)}{\left(\gamma +\Gamma_{\rm R}\right)^3 \left(2 \Gamma _{\rm L}+\Gamma _{\rm R}\right)^3}\nonumber\\
%&\stackrel{\Gamma_{\rm L}=\Gamma_{\rm R}}{\longrightarrow}&\frac{\gamma  \Gamma(7\gamma^2+10\gamma\Gamma+9\Gamma^2)}{27(\gamma +\Gamma)^3}\\
F_p&=&1-\frac{2 \gamma  \Gamma _{\rm L} \Gamma _{\rm R} \left(\gamma +2 \Gamma _{\rm L}+2 \Gamma _{\rm R}\right)}{\left(\gamma +\Gamma _{\rm R}\right)^2 \left(2 \Gamma _{\rm L}+\Gamma_{\rm R}\right)^2},
%\frac{4 \left(\gamma ^2+\Gamma _{\rm R} \left(\gamma +\Gamma _{\rm R}\right)\right) \Gamma _{\rm L}^2+2 \Gamma _{\rm R} \left(\gamma ^2+2 \Gamma _{\rm R} \left(\gamma+\Gamma _{\rm R}\right)\right) \Gamma _{\rm L}+\Gamma _{\rm R}^2 \left(\gamma +\Gamma _{\rm R}\right)^2}{\left(\gamma +\Gamma _{\rm R}\right)^2 \left(2 \Gamma_{\rm L}+\Gamma_{\rm R}\right)^2}
%\stackrel{\Gamma_{\rm L}=\Gamma_{\rm R}}{\longrightarrow}\frac{7\gamma^2+10\gamma\Gamma+9\Gamma^2}{9(\gamma+\Gamma)^2}.
\end{eqnarray}
which is mantained for all the low AC intensities regime, and the resonance fluorescence-like behaviour is completely lost, cf. Fig. \ref{hb}.

The sign of the electron-photon correlation depends on the asymmetry of the tunneling couplings. Concretely, in the case $\gamma\ll\Gamma_{\rm R}$, it is positive if $\Gamma_{\rm L}<\frac{\Gamma_{\rm R}^2}{2\gamma}$. 
%It may also be positive when $\Gamma_{\rm L}=\Gamma_{\rm L}=\Gamma$ if the photon emission rate satisfies $\gamma>\Gamma/5$:
%\begin{equation}
%r=\frac{\sqrt{\gamma}  \left(4 \gamma  \Gamma _{\rm L}^2-2 \Gamma _{\rm R}^2 \Gamma _{\rm L}+\Gamma _{\rm R}^2 \left(\gamma +\Gamma _{\rm R}\right)\right)}{\sqrt{2\left(\gamma +\Gamma _{\rm R}\right) \left(4 \Gamma _{\rm L}^2+\Gamma _{\rm R}^2\right) \left(4 \left(\gamma ^2+\Gamma _{\rm R} \gamma +\Gamma _{\rm R}^2\right) \Gamma   _{\rm L}^2+2 \Gamma _{\rm R} \left(\gamma ^2+2 \Gamma _{\rm R} \gamma +2 \Gamma _{\rm R}^2\right) \Gamma _{\rm L}+\Gamma _{\rm R}^2 \left(\gamma +\Gamma _{\rm R}\right)^2\right)}}.
%\end{equation}
Also if the photon emission rate is large enough compared to the tunneling rates. Concretely: once an electron occupies
the upper level, it will rather be relaxed to the lower level and tunnel to the collector than directly tunnel from
the upper level. Then, the probability of detecting consequently one photon and one electron increases, thus making the electron-photon correlation positive if:
\begin{equation}
\gamma>\frac{\Gamma_{\rm R}^2(2\Gamma_{\rm L}-\Gamma_{\rm R})}{4\Gamma_{\rm L}^2+\Gamma_{\rm R}^2}.
\end{equation}
This is more clearly seen when considering $\Gamma_{\rm L}=\Gamma_{\rm R}=\Gamma$:
%and the correlation coefficient (for $\Gamma_{\rm L}=\Gamma_{\rm R}=\Gamma$):
\begin{equation}\label{}
r=(5\gamma-\Gamma)\sqrt{\frac{\gamma}{10(\gamma+\Gamma)(7\gamma^2+10\gamma\Gamma+9\Gamma^2)}}.
\end{equation}
The coefficient:
\begin{equation}
\eta_p=1-\frac{2 \gamma  \left(7 \gamma ^3+41 \Gamma  \gamma ^2+52 \Gamma ^2 \gamma +36 \Gamma ^3\right)}{27 (\gamma +\Gamma )^4}
\end{equation}
also shows sub-Poissonian behaviour.

\subsection{High intensity limit: $\Omega\rightarrow\infty$}
For high AC field intensities, the contribution of the chemical potencial of the collector is only reflected in the occupation probabilities. Particularly important for the photonic dynamics is the probability of finding the QD in its empty and lower states, since it limits photon emission. The current, in this case, is:
\be
I_p=\frac{\gamma\Gamma_{\rm L}}{2\Gamma_{\rm L}+\Gamma_{\rm R}}. 
\ee
As seen in the previous regimes, the occupation of the empty state affects the sub-Poissonian statistics (expected for resonance fluorescence) by turning it to super-Poissonian values:
\begin{equation}\label{fphbinf}
F_p=1+\frac{\gamma  \Gamma _{\rm R}}{\left(2 \Gamma _{\rm L}+\Gamma _{\rm R}\right)^2}.
\end{equation}
Comparing to (\ref{fplbhi}) and (\ref{fpdcbhi}), the higher {\it unoccupation} of the QD involves a higher super-Poissonian character in the photonic statistics.

High intensities allow the emission of several photons before the electron is extracted to the collector. Also, an electron tunneling from the emitter to the upper level, can be extracted to the collector from the lower level without the emission of a photon. Then, the electron-photon correlation tends to be negative. However, if $\Gamma_{\rm L}$ is small, $\rho_0\approx1-2\Gamma_{\rm L}/\Gamma_{\rm R}\gg\rho_{1},\rho_2$, i. e., the probability of finding the QD empty is almost one. Then, the detection of photons and electrons
 is restricted to short lapses of time, which makes the electron-photon correlation positive:
\begin{equation}
r=\frac{\left(\Gamma _{\rm R}-2\Gamma _{\rm L}\right)\sqrt{\gamma\Gamma _{\rm R}}}{\sqrt{2\left(4 \Gamma _{\rm L}^2+\Gamma _{\rm R}^2\right) \left(4\Gamma _{\rm L}^2+4 \Gamma _{\rm R} \Gamma _{\rm L}+\Gamma _{\rm R} \left(\gamma +\Gamma _{\rm R}\right)\right)}}
%\stackrel{\Gamma_{\rm L}=\Gamma_{\rm R}}{\longrightarrow}-\sqrt{\frac{\gamma}{10(\gamma+9\Gamma)}}
\end{equation}

The third order coefficient gives:
\begin{equation}
\eta_p=1+\frac{3 \gamma  \Gamma _{\rm R} \left(8 \Gamma _{\rm L}^2-2 \left(\gamma -4 \Gamma _{\rm R}\right) \Gamma _{\rm L}+\Gamma _{\rm R} \left(\gamma +2 \Gamma _{\rm R}\right)\right)}{2\left(2 \Gamma _{\rm L}+\Gamma _{\rm R}\right)^4}.
\end{equation}

%+++++++++++++++++++++++++++++++++++++++++++++++++++++++++++++++++++++
\section{Selective tunneling configuration}
\label{secselect}

\begin{figure}[t]
\begin{center}
\includegraphics[width=2.25in,clip]{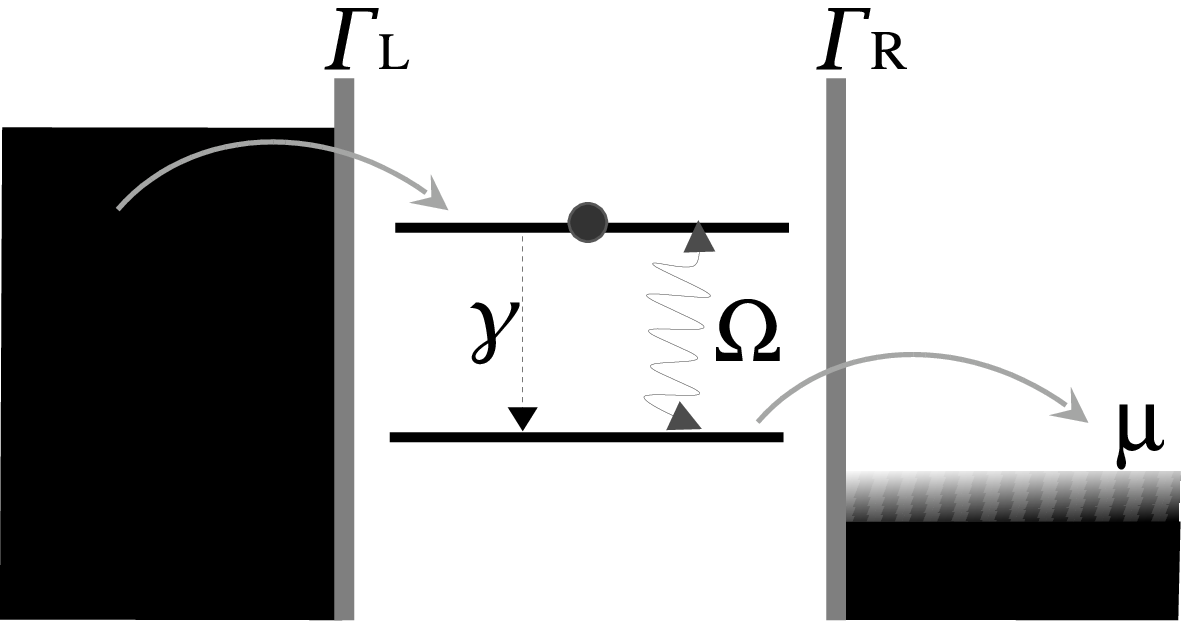}
\includegraphics[width=\linewidth,clip]{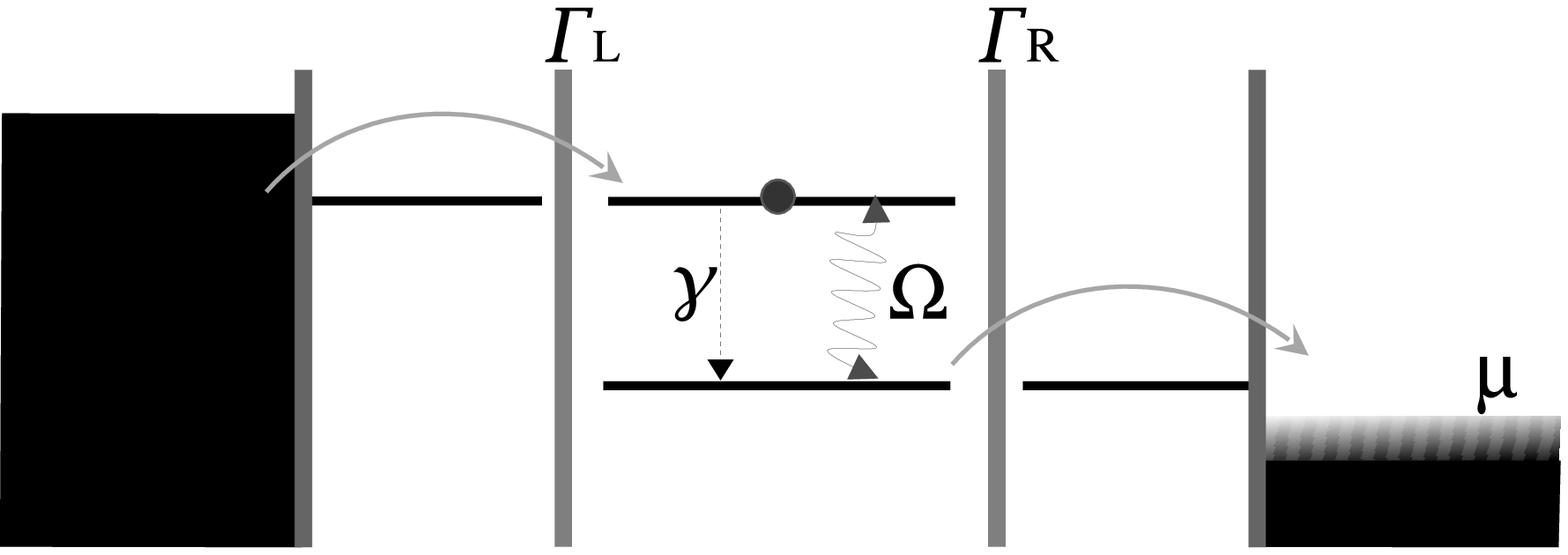}
\end{center}
\caption{\label{step}\small Diagrams of the {\it Selective tunneling} configuration, where each level is coupled to a different lead. ({\it Up}): the particular system considered here. ({\it Bottom}): a possible physical realization by considering a triple quantum dot where the lateral ones are strongly coupled to the leads so they behave as zero dimensional leads.}
\end{figure}
A particularly interesting configuration in the high bias regime ($f_1=f_2=0$) where the electron-photon correlation is paradigmatic, needs {\it unusual} coupling to the leads: electrons can enter only to the upper level and tunnel out only from the lower one. That is: $\Gamma_{2L}=\Gamma_{1R}=\Gamma$, $\Gamma_{1L}=\Gamma_{2R}=0$, cf. left diagram in Fig. \ref{step}.This selective coupling to the leads could be obtained by {\it zero-dimensional contacts} consisting in neighbour single-level QDs strongly coupled to the leads\cite{bryllert}. Then, if the level of the left(right) dot is resonant with the upper(lower) level, the emitter(collector) will be uncoupled of the lower(upper) level, see lower diagram in Fig. \ref{step}. Any eventual coherence between the central dot and the lateral ones is asumed to be rapidly damped by the coupling to the leads. We note here that such a system can also be used to modulate non-Markovian dynamics by tuning the strength of the coupling of the lateral dots to the leads. 

\subsection{Undriven case: Electron-photon identification}
In the absence of driving field, an electron that enters the upper
level can only be transferred to the collector after being relaxed
by the emission of one photon. Therefore, the electronic and
photonic statistics are completely identical and $c_{i0}=c_{0i}$. This configuration is analogue to having two single level quantum dots which are incoherently coupled, giving sub-Poissonian Fano factors\cite{kiesslich06} and maximal electron-photon correlation (see Appendix \ref{apstep}):
\begin{eqnarray}
I_e=I_p&=&\frac{\gamma\Gamma_{\rm L}\Gamma_{\rm R}}{\gamma\Gamma_{\rm L}+\gamma\Gamma_{\rm R}+\Gamma_{\rm L}\Gamma_{\rm R}}\\
\label{fephig}
F_e=F_p&=&1-\frac{2\gamma\Gamma_{\rm L}\Gamma_{\rm R}(\gamma+\Gamma_{\rm L}+\Gamma_{\rm R})}{(\gamma\Gamma_{\rm R}+\Gamma_{\rm L}(\gamma+\Gamma_{\rm R}))^2}\\
r&=&1.
\end{eqnarray}
The third cumulants give, for $\Gamma _{\rm L}=\Gamma _{\rm R}=\Gamma$:
\begin{equation}
\eta_e=\eta_p=
%1-6 \gamma  \Gamma _{\rm L} \Gamma _{\rm R}\frac{\left(\gamma ^2+\Gamma _{\rm R}^2\right) \Gamma _{\rm L}^3+\left(\gamma +\Gamma _{\rm R}\right) \left(\gamma ^2+\Gamma_{\rm R} \left(\gamma +\Gamma _{\rm R}\right)\right) \Gamma _{\rm L}^2+2 \gamma ^2 \Gamma _{\rm R}^2 \Gamma _{\rm L}+\gamma ^2 \Gamma _{\rm R}^2 \left(\gamma +\Gamma_{\rm R}\right)}{\left(\gamma  \Gamma _{\rm R}+\Gamma _{\rm L} \left(\gamma +\Gamma _{\rm R}\right)\right)^4}
1-6\gamma\frac{\left(3\gamma^2+\Gamma^2\right)\Gamma+\left(\gamma +\Gamma\right)\left(2\gamma^2+\Gamma\left(\gamma +\Gamma\right)\right)}{\left(2\gamma +\Gamma\right)^4}
\end{equation}

\subsection{Driven case}

\begin{figure}[t]
\begin{center}
\includegraphics[width=\linewidth,clip]{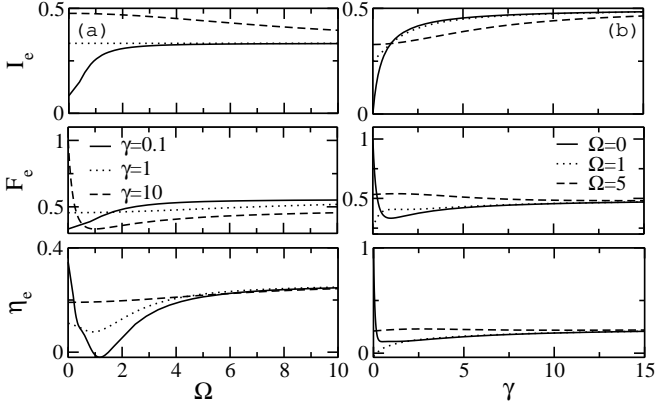}
\end{center}
\caption{\label{eZZW}\small {\bf Selective tunneling:} Electronic current, Fano factor and skewness as a function of (a) the driving intensity for different photon emission rates and (b) the photon emission rate for different field intensities. $\Gamma_{\rm L}=\Gamma_{\rm R}=\Gamma=1$.
}
\end{figure}
\begin{figure}[t!]
\begin{center}
\includegraphics[width=\linewidth,clip]{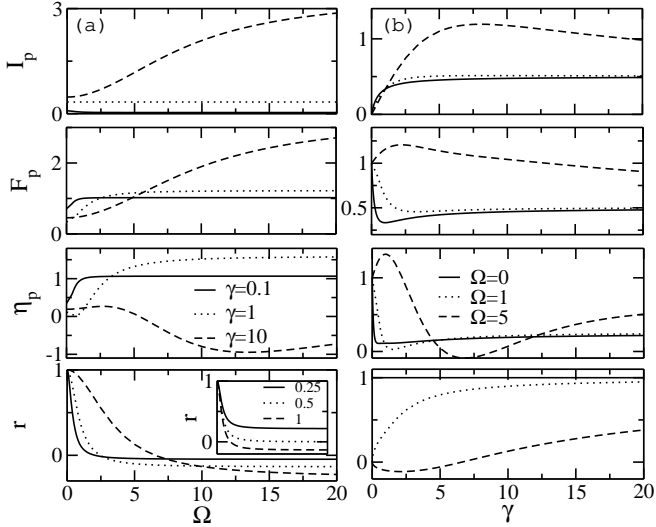}
\end{center}
\caption{\label{phZZW}\small {\bf Selective tunneling:} Photonic current, Fano factor and skewness as a function of (a) the driving intensity, for different photon emission rates, and (b) the photon emission rate, for different field intensities. $\Gamma_{\rm L}=\Gamma_{\rm R}=\Gamma=1$. Inset: Electron-photon correlation coefficient as a function of the field intensity for different couplings to the left contact, $\Gamma_{\rm L}$, with $\Gamma_{\rm R}=\gamma=1$. It is possible to tune the sign of the electron-photon correlation by means of the tunneling coupling asymmetry.
}
\end{figure}

The ac field allows the tunneling of an electron to the collector
without having previously emitted a photon as well as the emission
of several photons from the relaxation of the same electron. This makes the electronic and photonic currents differ, thus uncorrelating the electronic and photonic statistics. More interestingly, by looking at the dependence of the electronic and photonic currents with the detuning:
\begin{widetext}
\begin{equation}
I_e=\frac{\Gamma _{\rm L} \Gamma _{\rm R} \left(4 \gamma  \Delta _{\omega }^2+\left(\gamma +\Gamma _{\rm R}\right) \left(\gamma ^2+\Gamma _{\rm R} \gamma +\Omega^2\right)\right)}{\Gamma _{\rm L} \left(\gamma +\Gamma _{\rm R}\right) \left(\gamma ^2+2 \Gamma _{\rm R} \gamma +2 \Omega ^2+\Gamma _{\rm R}^2+4 \Delta _{\omega}^2\right)+\Gamma _{\rm R} \left(4 \gamma  \Delta _{\omega }^2+\left(\gamma +\Gamma _{\rm R}\right) \left(\gamma ^2+\Gamma _{\rm R} \gamma +\Omega^2\right)\right)}
\end{equation}
\begin{equation}
I_p=\frac{\gamma  \Gamma _{\rm L} \left(4 \Gamma _{\rm R} \Delta _{\omega }^2+\left(\gamma +\Gamma _{\rm R}\right) \left(\Omega ^2+\Gamma _{\rm R} \left(\gamma +\Gamma
   _{\rm R}\right)\right)\right)}{\Gamma _{\rm L} \left(\gamma +\Gamma _{\rm R}\right) \left(\gamma ^2+2 \Gamma _{\rm R} \gamma +2 \Omega ^2+\Gamma _{\rm R}^2+4 \Delta
   _{\omega }^2\right)+\Gamma _{\rm R} \left(4 \gamma  \Delta _{\omega }^2+\left(\gamma +\Gamma _{\rm R}\right) \left(\gamma ^2+\Gamma _{\rm R} \gamma +\Omega
   ^2\right)\right)},
\end{equation}
%\end{widetext}
it can be seen that their second order derivative obeys:
\begin{equation}
\label{partIep}
\frac{\partial^2 I_{e(p)}}{\partial \Delta_\omega^2}\propto \Gamma_{\rm R}-\gamma,
\end{equation}
which is reflected in a resonance to anti-resonance crossover. As a consequence, one can extract information on the sample-depending spontaneous photon emission rate, $\gamma$, by externally modifying the tunneling couplings to the collector\cite{genovaRF}. The system, in this case, behaves as a {\it photon emission rate probe}.

%\bea
%I_e&=&\frac{\Gamma _{\rm L} \Gamma _{\rm R} \left(\gamma^2+\Gamma _{\rm R} \gamma +\Omega ^2\right)}{\Gamma _{\rm R}\left(\gamma ^2+\Gamma _{\rm R} \gamma +\Omega^2\right)+\Gamma _{\rm L} \left(\gamma ^2+2 \Gamma _{\rm R} \gamma +2 \Omega ^2+\Gamma _{\rm R}^2\right)}\\
%\frac{\Gamma  \left(\gamma ^2+\Gamma  \gamma +\Omega ^2\right)}{2 \gamma ^2+3 \Gamma  \gamma +\Gamma ^2+3 \Omega ^2}\\
%I_p&=&\frac{\gamma  \Gamma _{\rm L} \left(\Omega ^2+\Gamma_{\rm R} \left(\gamma +\Gamma _{\rm R}\right)\right)}{\Gamma _{\rm R} \left(\gamma ^2+\Gamma _{\rm R} \gamma +\Omega^2\right)+\Gamma _{\rm L} \left(\gamma ^2+2 \Gamma _{\rm R} \gamma +2 \Omega ^2+\Gamma _{\rm R}^2\right)}
%\eea
%and un-correlates the electronic and photonic statistics. 
Considering $\Gamma_{\rm L}=\Gamma_{\rm R}=\Gamma$, for simplicity, and the resonance condition, $\Delta_\omega=0$, we obtain for the Fano factors:
\begin{eqnarray}
F_e&=&1-\frac{2 \left(\gamma ^4+4 \Gamma  \gamma ^3+5 \Gamma ^2 \gamma ^2+3 \Omega ^2 \gamma ^2+2 \Gamma ^3 \gamma +3 \Gamma  \Omega ^2 \gamma +2
   \Omega ^4+4 \Gamma ^2 \Omega ^2\right)}{\left(2 \gamma ^2+3 \Gamma  \gamma +\Gamma ^2+3 \Omega ^2\right)^2}\\
F_p&=&1-\frac{2 \gamma  \left(2 \Gamma ^4-\Omega ^2 \Gamma ^2+\gamma ^3 \Gamma -\Omega ^4+\gamma ^2 \left(4 \Gamma ^2-\Omega ^2\right)+\gamma  \left(5
   \Gamma ^3+6 \Omega ^2 \Gamma \right)\right)}{\Gamma  \left(2 \gamma ^2+3 \Gamma  \gamma +\Gamma ^2+3 \Omega ^2\right)^2},
\end{eqnarray}
\end{widetext}
cf. Figs. \ref{eZZW} and \ref{phZZW}. From the Fano factor, it can be seen that the electrons obey
sub-Poissonian statistics while the photons become
super-Poissonian for high enough field intensities. The driving
field also contributes to make the electron-photon correlation
coefficient negative, see Appendix \ref{apstep}.

%\begin{eqnarray}
%&r&=\sqrt{\gamma}\left(\Gamma(\gamma +\Gamma )^3 \left(2 \gamma ^2+\Gamma ^2\right)-\gamma\Gamma(\gamma -11 \Gamma )(\gamma+\Gamma ) \Omega ^2-\left(\gamma ^2+\Gamma  \gamma -4 \Gamma ^2\right) \Omega ^4-\Omega ^6\right)\nonumber\\
%&&\times\large[\left((2 \gamma +9 \Gamma ) \Omega ^4+2 \left(\gamma ^3+10 \Gamma ^2\gamma +3 \Gamma ^3\right) \Omega ^2+\Gamma  (\gamma +\Gamma )^2 \left(2 \gamma ^2+\Gamma ^2\right)\right)\\
%&&\times \left(\Omega ^2+\gamma  (\gamma+\Gamma )\right) \left(\Omega ^2+\Gamma  (\gamma +\Gamma )\right) \left(5 \Omega ^4+2 \left(3 \gamma ^2+6 \Gamma  \gamma -\Gamma ^2\right)\Omega ^2+(\gamma +\Gamma )^2 \left(2 \gamma ^2+\Gamma ^2\right)\right)\large]^{-1/2}.\nonumber
%\end{eqnarray}

\begin{figure}[t]
\begin{center}
\includegraphics[width=\linewidth,clip]{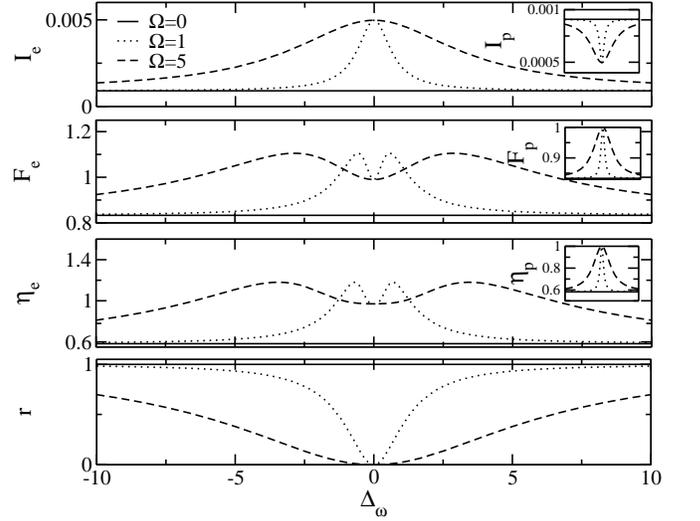}
\end{center}
\caption{\label{ephZZasimNorel}\small {\bf Selective tunneling:} Electronic current, Fano factor, skewness and electron-photon correlation as functions of the frequency detuning for different field intensities. $\Gamma_{\rm L}=1$, $\Gamma_{\rm R}=0.001$ and $\gamma=0.001$. In the insets the corresponding photonic values are shown. When the photon emission rate is much smaller than the tunneling rates and $\Gamma_{\rm L}>\Gamma_{\rm R}$, the system behaves as a coherently coupled double quantum dot, showing a sub-Poissonian minimum in the Fano factor which is between two super-Poissonian peaks. 
%Since only relaxation processes are considered here, these two peaks are symetric. 
}
\end{figure}

In the absence of relaxation, this configuration can be mapped into a coherently coupled single level double quantum dot, where interdot hopping played the role of the ac driving (whithin the rotating wave approximation). 
%Then, as seen in Refs. \cite{kiesslichHaug,yoHaug}, it is interesting to consider the asymetric coupling case, $\Gamma_{\rm R}<\Gamma_{\rm L}$. The effect of the bias voltage there is similar to tuning the ac frequency in the present case.  
Then, in the particular case where $\Gamma_{\rm R}<\Gamma_{\rm L}$, the noise is sub-Poissonian in resonance, having two super-Poissonian peaks in its vicinity, when the influence of photons is small,  as seen in Fig. \ref{ephZZasimNorel}. For $\gamma=0$\cite{elattari}:
\bea
F_e&=&1\nonumber\\[-2.5mm]
&&\\[-2.5mm]
&&\!\!\!\!\!\!\!\!\!\!\!\!-\frac{2 \Omega ^2 \Gamma_{\rm L} \left(4 \left(\Gamma _{\rm R}-\Gamma _{\rm L}\right) \Delta _{\omega }^2+\Gamma _{\rm R} \left(2 \Omega ^2+\Gamma _{\rm R}^2+3 \Gamma _{\rm L}\Gamma _{\rm R}\right)\right)}{\left(\Gamma _{\rm R} \Omega ^2+\Gamma _{\rm L} \left(2 \Omega ^2+\Gamma _{\rm R}^2+4 \Delta _{\omega }^2\right)\right)^2}.\nonumber 
\eea
This kind of features has been the subject of recent works in double quantum dot systems where the double peak structure in the electronic Fano factor becomes asymmetric by the effect of temperature\cite{kiesslichHaug,yoHaug}.
In our case, the level energies are not shifted, so the contribution of photon emission is constant all over the ac frequency tuning and the double peak remains symmetric. As a consequence, transport is not quenched by detuning and electronic noise is sub-Poissonian far from resonance (where the ac field has no effect on transport), as expected from (\ref{fephig}). Interestingly, the maximal electron-photon correlation observed far from resonance vanishes for $\Delta_\omega=0$.  

\begin{figure}[t]
\begin{center}
\includegraphics[width=\linewidth,clip]{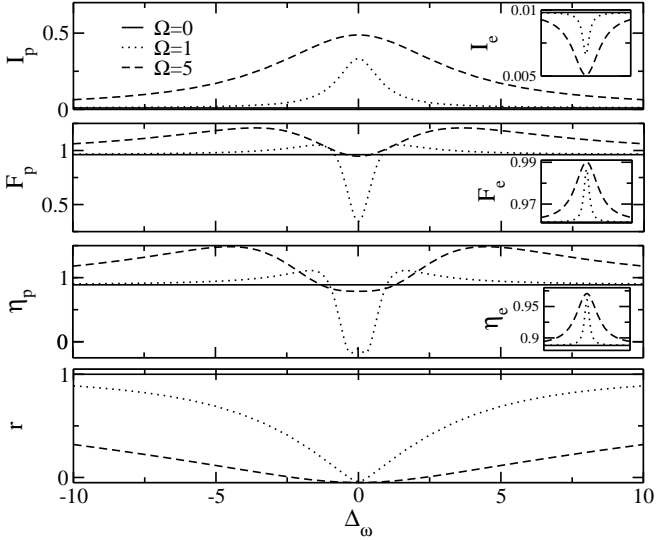}
\end{center}
\caption{\label{pheZZasimRel}\small {\bf Selective tunneling:} Photonic current, Fano factor, skewness and electron-photon correlation as functions of the frequency detuning for different field intensities. $\Gamma_{\rm L}=\gamma=1$ and $\Gamma_{\rm R}=0.001$. The insets show the electronic correspondents. The double peak in the electronic Fano factor seen in Fig. \ref{ephZZasimNorel} is washed out by a larger photon emission rate. 
}
\end{figure}

The double peak in the electronic Fano factor is washed out for larger photon emission rates, even for the case $\Gamma_{\rm L}\gg\Gamma_{\rm R}$:
\bea
F_e=1-\frac{2 \Gamma_{\rm R}}{\gamma } \left(1-\frac{\left(5 \gamma ^2+4 \Omega ^2+4 \Delta _{\omega }^2\right) \Omega ^2}{\left(\gamma ^2+2 \Omega ^2+4 \Delta _{\omega}^2\right)^2}\right.\nonumber\\[-2.5mm]
\\[-2.5mm]
\left.+\frac{\gamma}{\Gamma _{\rm L}}\left(1-\frac{\Omega ^2}{\gamma ^2+2 \Omega ^2+4 \Delta _{\omega }^2}\right)\right)+O\left(\Gamma _{\rm R}^2\right),\nonumber
\eea
as seen in the insets of Fig. \ref{pheZZasimRel}. On contrary, in this regime, it is the photonic noise which is sub-Poissonian but for two super-Poissonian regions around the resonant frequency, recovering the resonance fluorescence behaviour, see Fig. \ref{pheZZasimRel}.

\subsection{High intensity limit: $\Omega\rightarrow\infty$}
An intense driving involves the delocalization of the electron between the upper and lower level, so it has the same probability of occupying each of them: $\rho_1=\rho_2=\Gamma_{\rm L}/(2\Gamma_{\rm L}+\Gamma_{\rm R})$. In this case, the resonant currents are:
\be
\frac{I_e}{\Gamma_{\rm L}\Gamma_{\rm R}}=\frac{I_p}{\gamma\Gamma_{\rm L}}=\frac{1}{2\Gamma_{\rm L}+\Gamma_{\rm R}}.
\ee
The electronic dynamics becomes independent from photon emission so the Fano factor coincides with that obtained for transport through a double quantum dot in the absense of dissipation, being sub-Poissonian at resonance ($\Delta_\omega=0$):
\begin{equation}
F_e=1-\frac{4\Gamma_{\rm L}\Gamma_{\rm R}}{(2\Gamma_{\rm L}+\Gamma_{\rm R})^2}
\end{equation}
and super-Poissonian close to resonance, if $\Gamma_{\rm L}>\Gamma_{\rm R}$\cite{kiesslichHaug,yoHaug}.
 
The high probability of finding the QD empty, $\rho_0=\Gamma_{\rm R}/(2\Gamma_{\rm L}+\Gamma_{\rm R})$, kills the {\it resonance fluorescence-like} photon anti-bunching and the photonic statistics become super-Poissonian:
\begin{equation}
F_p=1+\frac{2\gamma\Gamma_{\rm R}}{(2\Gamma_{\rm L}+\Gamma_{\rm R})^2}.
\end{equation}
As discussed in previous sections, the electron-photon correlation is lost by the influence of the ac field. However, if the coupling to the leads is asymmetric and $\Gamma_{\rm R}>2\Gamma_{\rm L}$, once the electron has tunneled out to the collector, the QD remains empty for a long period of time (compared to the lapse of time that it spends occupied). Then, the detection of electrons and photons is restricted to the short periods of time, so $r$ remains positive, cf. inset in Fig. \ref{phZZW}a:
\begin{equation}
r=\frac{\sqrt{\gamma\Gamma _{\rm R}}\left(\Gamma _{\rm R}-2 \Gamma _{\rm L}\right)}{\sqrt{\left(4 \Gamma _{\rm L}^2+\Gamma _{\rm R}^2\right) \left(4 \Gamma_{\rm L}^2+4 \Gamma _{\rm R} \Gamma _{\rm L}+\Gamma _{\rm R} \left(2 \gamma +\Gamma _{\rm R}\right)\right)}}.
\end{equation}
The third order cumulants give:
\be
\eta_e=1-\frac{12 \Gamma _{\rm L} \Gamma _{\rm R} \left(4 \Gamma _{\rm L}^2+\Gamma _{\rm R}^2\right)}{\left(2 \Gamma _{\rm L}+\Gamma _{\rm R}\right)^4},
\ee
for electrons, and:
\be
\eta_p=1+\frac{6 \gamma  \Gamma _{\rm R} \left(4 \Gamma _{\rm L}^2-2 \left(\gamma -2 \Gamma _{\rm R}\right) \Gamma _{\rm L}+\Gamma _{\rm R} \left(\gamma +\Gamma_{\rm R}\right)\right)}{\left(2 \Gamma _{\rm L}+\Gamma _{\rm R}\right)^4},
\ee
for photons.

%*****************************************************
%*****************************************************
%*****************************************************
\section{Level-dependent tunneling}
\label{secleveldep}

If the left and right barriers are equal, the tunneling events may differ depending on which level participates. This can be due to the concrete orbital distribution of each level. Then, one has $V_{{\rm L}i}=V_{{\rm R}i}$ for the couplings in (\ref{hamRF}) and electronic transport can be parametrized, if both level are within the transport window, by the tunneling rates $\Gamma_{2}=2\pi d_l|V_{{l}2}|^2$ and $\Gamma_{1}=2\pi d_l|V_{{l}1}|^2$ when the electron tunnels to or from the upper or the lower level, respectively, through any barrier $l$\cite{lambertTh}.

The equations of motion for the generating function, $\dot G(t,s_e,s_{p})=M(s_e,s_{p})G(t,s_e,s_{p})$, and the density matrix, $\dot \rho(t)=M(1,1)\rho(t)$ (after setting $s_e=s_p=1$), are then given by the matrix:
\begin{widetext}
\begin{equation}
\displaystyle
M(s_e,s_{p})=\left(\begin{array}{ccccc}
-\Gamma_{2}-\Gamma_{1} & s_e\Gamma_{1} & 0 & 0 & s_e\Gamma_{2} \\
\Gamma_{1} &  -\Gamma_{1} & i\frac{\Omega}{2} & -i\frac{\Omega}{2} & s_{p}\gamma  \\
0 & i\frac{\Omega}{2} & \Lambda_{12}+i\Delta_{\omega} & 0 & -i\frac{\Omega}{2} \\
0 & -i\frac{\Omega}{2} & 0 &\Lambda_{12}-i\Delta_{\omega} & i\frac{\Omega}{2} \\
\Gamma_{2} & 0 & -i\frac{\Omega}{2} & i\frac{\Omega}{2} & -\gamma-\Gamma_{2}\\
\end{array}  \right)
\end{equation}
in the same matrix form chosen to write (\ref{MLR}). In this case, the decoherence term is given by: $\Lambda_{12}=-\frac{\Gamma_{2}+\Gamma_{1}+\gamma}{2}$.

%\begin{figure}[t]
%\begin{center}
%\includegraphics[width=\linewidth,clip]{phLevelsud.eps}
%\end{center}
%\caption{\label{udgW}\small {\bf Level dependent tunneling:} Photonic current, Fano factor and skewness and electronb-photon correlation coefficient as a function of (a) the field intensity, for different photon emission rates, and (b) the photon emission rate, for different field intensities. $\Gamma_{1}=\Gamma_{2}=1$. The electronic statistics is independent of both the electric field and relaxation.
%The field introduces negative correlation. $\tilde\Omega=\Omega/\Gamma$.
%}
%\end{figure}
The dependence on the level which is occupied introduces the effect of the driving field and photon emission in the electronic current even in the high bias regime. If, for instance, $\Gamma_1<\Gamma_2$, photon emission photon emission will contribute to decrease the flow of electrons.
The electronic and photonic currents are, in the general case:

\begin{eqnarray}
I_e&=&\frac{\left(\Gamma _{2}+\Gamma _{1}\right) \left(\left(\Gamma_{2}+\Gamma_{1}\right)\left(\Gamma_{2}\Gamma_{1}+\Omega ^2\right)+\gamma\Gamma_{1}\left(\gamma+\Gamma_{1}+2\Gamma_{2}\right)\right)}{\left(\gamma +3 \Gamma _{1}\right) \Gamma _{2}^2+\left(\gamma ^2+3 \Omega ^2+3 \Gamma _{1} \left(2\gamma +\Gamma _{1}\right)\right) \Gamma _{2}+\Gamma _{1} \left(2 \gamma ^2+2 \Gamma _{1} \gamma +3 \Omega ^2\right)}\\
I_p&=&\frac{\gamma\left(\left(\Gamma_{2}+\Gamma_{1}\right)\left(\Gamma_{2}\Gamma_{1}+\Omega ^2\right)+\gamma\Gamma_{2}\right)}{\left(\gamma +3 \Gamma _{1}\right) \Gamma _{2}^2+\left(\gamma ^2+3 \Omega ^2+3 \Gamma _{1} \left(2 \gamma +\Gamma_{1}\right)\right) \Gamma _{2}+\Gamma _{1} \left(2 \gamma ^2+2 \Gamma _{1} \gamma +3 \Omega ^2\right)}.
\end{eqnarray}
\end{widetext}
As expected, if $\Gamma_{1}<\Gamma_{2}$, the emission of photons inhibits electronic transport. However, the opposite is not true: if $\Gamma_{2}<\Gamma_{1}$, electrons will rather tunnel through the lower level, thus avoiding photon emission. These two limiting cases will be further analyzed below. In the case, $\Gamma_1=\Gamma_2$, the electronic current is independent of both the relaxation rate and the driving intensity, recovering the behaviour described in section \ref{sechb}, cf. Fig. \ref{hb}.

\subsection{Undriven case.}

In the absence of the driving field, the electronic and photonic currents are given by:
\be
\frac{I_e}{\Gamma_1(\gamma+\Gamma_2)(\Gamma_1+\Gamma_2)}=\frac{I_p}{\gamma\Gamma_1\Gamma_2}=\frac{1}{2\gamma\Gamma_1+(\gamma+3\Gamma_1)\Gamma_2}.
\ee
In this case, the difference in the tunneling rates of each level is enough to define the sub- or super-Poissonian electronic statistics: % in an unblocked version of dynamical channel blockade:
%photon emission may affect dramatically the transport characteristics, as discussed above, favouring the occupation of the lower level.
%\be
%F_e=1+\frac{2 \Gamma _{2}\left(\Gamma _{1}^3+(\gamma+\Gamma_{1})\Gamma_{2}^2\right)-2\Gamma_{1}\left(\Gamma_1(\gamma+2\Gamma_2)^2 +\gamma\Gamma_{2}(\gamma+\Gamma_{2})\right)}{\left(2 \gamma  \Gamma _{1}+\left(\gamma +3 \Gamma _{1}\right) \Gamma _{2}\right)^2}.
%\ee
\be
F_e=1+\frac{2 \Gamma _{2}\left(\Gamma _{1}^3+\tilde\Gamma_{1}\Gamma_{2}^2\right)-2\Gamma_{1}\left(\Gamma_1(\gamma+2\Gamma_2)^2 +\gamma\Gamma_{2}\tilde\Gamma_{2}\right)}{\left(2 \gamma  \Gamma _{1}+\left(\gamma +3 \Gamma _{1}\right) \Gamma _{2}\right)^2},
\ee
where we have called $\tilde\Gamma_i=\gamma+\Gamma_i$.
Interestingly, in the absence of photon relaxation, the Fano factor depends linearly on the asymmetry and increases as one of the levels becomes uncoupled of the leads (this case will be considered below):
\begin{equation}
F_e(\gamma\ll\Gamma_i)=1+\frac{2}{9}\left(\frac{\Gamma_{1}}{\Gamma_{2}}+\frac{\Gamma_{2}}{\Gamma_{1}}-4\right).
\end{equation}
Photon emission diminish this effect by contributing to make the electrons be extracted from the lower level. Then, the 
sub-Poissonian shot noise observed in the high bias regime, cf. Eq. (\ref{fehb}), is recovered:
\begin{equation}
F_e(\gamma\gg\Gamma_i)=1-\frac{2\Gamma_{1}(\Gamma_{1}+\Gamma_{2})}{\left(2\Gamma_{1}+\Gamma_{2}\right)^2}.
\end{equation}

On the other hand, photonic statistics remain sub-Poissonian, independently of the configuration:
\begin{equation}
F_p=1-\frac{2 \gamma  \Gamma _{1} \Gamma _{2} \left(\gamma +2 \Gamma _{1}+2 \Gamma _{2}\right)}{\left(2 \gamma  \Gamma _{1}+\left(\gamma +3 \Gamma _{1}\right)\Gamma _{2}\right)^2}.
\end{equation}
The electron-photon correlation coefficient (see Appendix \ref{apseltun})
shows how electrons and photons can be uncorrelated by the manipulation of the tunneling rates. 

\subsection{High intensity limit: $\Omega\rightarrow\infty$}
As the driving field couples the two levels, it tends to annihilate the particular behaviour introduced by the different couplings to the leads. Thus, the currents depend simply on their correnspondent rate:
\be
\frac{I_e}{\Gamma_1+\Gamma_2}=\frac{I_p}{\gamma}=\frac{1}{3},
\ee
while and the electronic Fano factor and skewness become independent of the tunneling couplings:
%\begin{eqnarray}
$F_e=\frac{5}{9}$ and $\eta_e=\frac{7}{27}$,
%\end{eqnarray}
consistently with (\ref{fehb}). 
That is not the case for the photonic statistics, whose second and third moments depend on the rates:
\begin{eqnarray}
F_p&=&1+\frac{2 \gamma}{9\left(\Gamma _{1}+\Gamma _{2}\right)}\\
\eta_p&=&1-\frac{2 \gamma  \left(\gamma -9 \Gamma _{1}-9 \Gamma _{2}\right)}{27 \left(\Gamma _{1}+\Gamma _{2}\right)^2}.
\end{eqnarray}
As expected, electron-photon correlation becomes negative:
\begin{equation}
r=-\sqrt{\frac{\gamma}{5\left(2\gamma+9\left(\Gamma _{1}+\Gamma _{2}\right)\right)}}.
\end{equation}

\subsection{$\Gamma_{1}\ll\Gamma_{2}$ limit}
\begin{figure}[t]
\begin{center}
\includegraphics[width=\linewidth,clip]{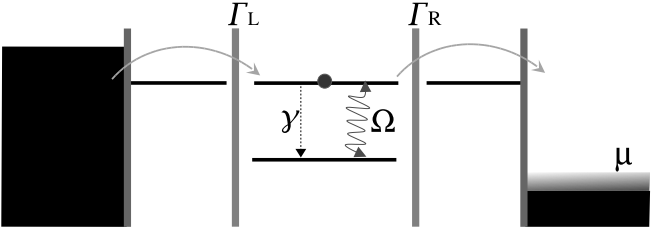}
\end{center}
\caption{\label{hbu}\small Schematic diagram of the proposed setup for a level dependent tunneling configuration where $\Gamma_{1}\ll\Gamma_{2}$.}
\end{figure}

The zero-dimensional contacts introduced in the previous section can also be employed to simulate energy-dependent tunneling. If both zero-dimensional contacts are aligned (by tuning the gate voltages of the left and right QDs) with the same level of the QD, transport through the other level will be strongly suppressed, cf. Fig. \ref{hbu}. Thus, the occupation of the off resonant level blocks the electronic current. %This configuration can be analogue to double quantum dot systems where only one of them is coupled to the leads that have been proposed as qubits.

If the levels of the surrounding QDs are aligned with the upper level, in the absence of driving, as soon as the lower level is occupied (by the relaxation of an electron from the upper level), transport is canceled in a high bias version of dynamical channel blockade. Thus, electrons flow in bunches, while photonic transport is highly supressed.

\begin{figure}[t]
\begin{center}
\includegraphics[width=\linewidth,clip]{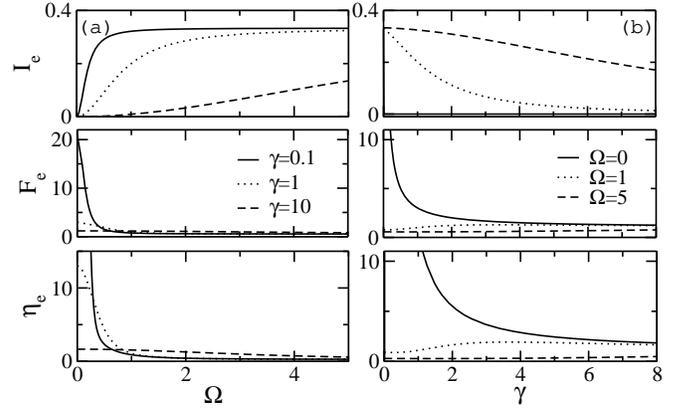}
\end{center}
\caption{\label{udWe}\small {\bf Level dependent tunneling:} Electronic current, Fano factor and skewness as a function of (a) the field intensity, for different photon emission rates, and (b) the photon emission rate for different field intensities, for the case: $\Gamma_{1}=10^{-5}$ and $\Gamma_{2}=1$.
}
\end{figure}

\begin{figure}[t]
\begin{center}
\includegraphics[width=\linewidth,clip]{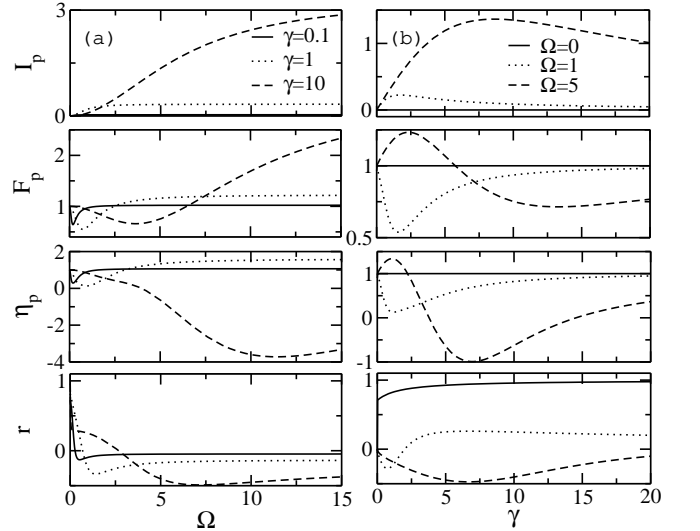}
\end{center}
\caption{\label{udWph}\small {\bf Level dependent tunneling:} Photonic current, Fano factor and skewness and electron-photon correlation coefficient as a function of (a) the field intensity, for different photon emission rates, and (b) the photon emission rate for different field intensities, for the case: $\Gamma_{1}=10^{-5}$ and $\Gamma_{2}=1$.
}
\end{figure}

Again, the driving field removes the blockade, producing finite electronic and photonic currents:
\begin{equation}
\frac{I_e}{\Gamma_{2}}=\frac{I_p}{\gamma}=\frac{\Omega^2}{\gamma^2+\gamma\Gamma_{2}+3\Omega^2}
\end{equation}
thus reducing the super-Poissonian electron noise:
\begin{equation}
F_e=1+2\frac{\gamma\Gamma_2(\gamma+\Gamma_{2})^2-\Omega^2(\gamma^2+2\gamma\Gamma_2-\Gamma_2^2)-2\Omega^4}{(\gamma(\gamma+\Gamma_{2})+3\Omega^2)^2},
%F_e=1+2\frac{\gamma(\gamma+2\Gamma_{2})\left(\Gamma_{2}\gamma-\Omega^2\right)+\Gamma_2^2(2\gamma\Gamma_2+\Omega^2)-2\Omega^4}{(\gamma^2+\gamma\Gamma_{2}+3\Omega^2)^2},
\end{equation}
which becomes sub-Poissonian for high enough driving intensities, cf. Fig. \ref{udWe}. This configuration resembles a single level quantum dot coupled to a localized state, where super-Poissonian shot noise has been predicted\cite{djuricT}. The obtained Fano factor recovers their result for a non-dissipative situation, $\gamma=0$. Similar models were proposed to explain enhanced shot noise in single quantum dots\cite{safonov}. For low intensities, the photonic noise is sub-Poissonian, resembling the resonance fluorescence but, for $\Omega>\sqrt{2\Gamma_{2}(2\gamma+\Gamma_{2})}$, the contribution of the empty state turns it super-Poissonian, cf. Fig. \ref{udWph}:
\begin{equation}
F_p=1-2 \gamma\Omega^2\frac{2\Gamma_{2}(2\gamma+\Gamma_{2})-\Omega^2}{\Gamma_{2}\left(\gamma^2+3\Omega^2+\gamma\Gamma_{2}\right)^2}.
\end{equation}
It is interesting to note that, though the electronic and photonic mean counts are proportional, their variances are not, which is reflected in the electron-photon correlation
%\be
%r=\frac{\gamma  \left(\gamma  \Gamma _u^3+\left(2 \gamma ^2-\Omega ^2\right) \Gamma _u^2+\left(\gamma ^3-6 \gamma  \Omega ^2\right) \Gamma_u-\Omega ^2 \left(\gamma ^2+\Omega ^2\right)\right)}{\sqrt{\gamma  \left(\gamma ^4+4 \Omega ^2 \gamma ^2+2 \Gamma _u^3 \gamma +2 \left(2 \gamma ^2+\Omega ^2\right) \Gamma _u \gamma +5 \Omega ^4+\left(5 \gamma ^2+2 \Omega ^2\right) \Gamma _u^2\right) \left(2 \gamma  \Omega^4+\gamma ^2 \Gamma _u^3+2 \gamma  \left(\gamma ^2+\Omega ^2\right) \Gamma _u^2+\left(\gamma ^4-2 \Omega ^2 \gamma ^2+9 \Omega ^4\right)\Gamma _u\right)}}
%\ee
%\begin{equation}
%c_{11}=\gamma\Omega^2\frac{\gamma\Gamma_{2}(\gamma+\Gamma_{2})^2-\Omega^2(\gamma^2+\Gamma_{2}^2+6\gamma\Gamma_{2})-\Omega^4}{(\gamma^2+3\Omega^2+\gamma\Gamma_{2})^3}
%\end{equation}
that gives $r<1$, see (\ref{c11G1_0}). 

The undriven case gives a Fano factor that diverges when the photon emission is reduced, $F_e=1+2\frac{\Gamma_2}{\gamma}$, and $\eta_e=\frac{\gamma^2+6\Gamma_{2}^2(\gamma+\Gamma_{2})}{\gamma^2}$. In this case, relaxation becomes an stochastic process.

\subsection{$\Gamma_{2}\ll\Gamma_{1}$ limit}
\begin{figure}[t]
\begin{center}
\includegraphics[width=\linewidth,clip]{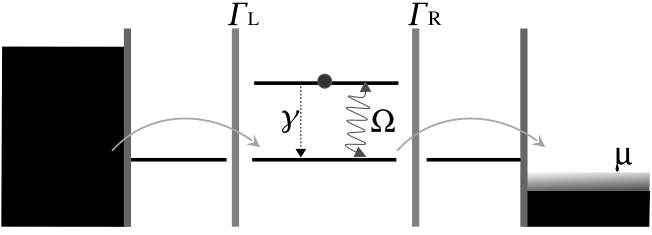}
\end{center}
\caption{\label{hbd}\small Schematic diagram of the proposed setup for a level dependent tunneling configuration where $\Gamma_{2}\ll\Gamma_{1}$.}
\end{figure}

This case is similar to the previous one with the difference that the contribution of photon emission has the opposite effect: the upper level is very weakly coupled to the leads so its population quenches the electronic current, cf. Fig. \ref{hbd}. Therefore, relaxation by photon emission contributes to unblock the electronic transport.

\begin{figure}[t]
\begin{center}
\includegraphics[width=\linewidth,clip]{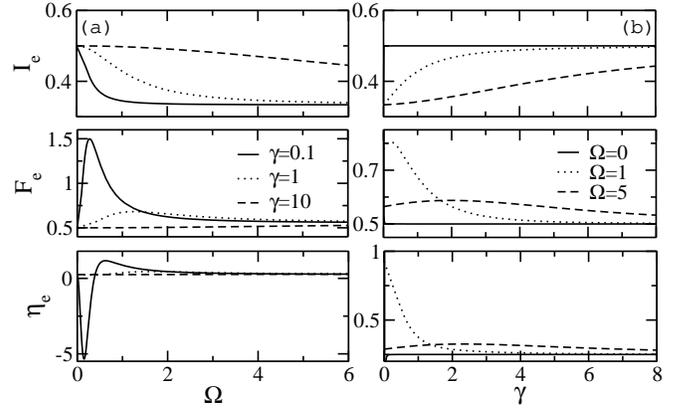}
\end{center}
\caption{\label{eudW}\small {\bf Level dependent tunneling:} Electronic current, Fano factor and skewness as a function of (a) the field intensity, for different photon emission rates, and (b) the photon emission rate for different field intensities, for the case: $\Gamma_{1}=1$ and $\Gamma_{2}=10^{-5}$.
}
\end{figure}

\begin{figure}[t]
\begin{center}
\includegraphics[width=\linewidth,clip]{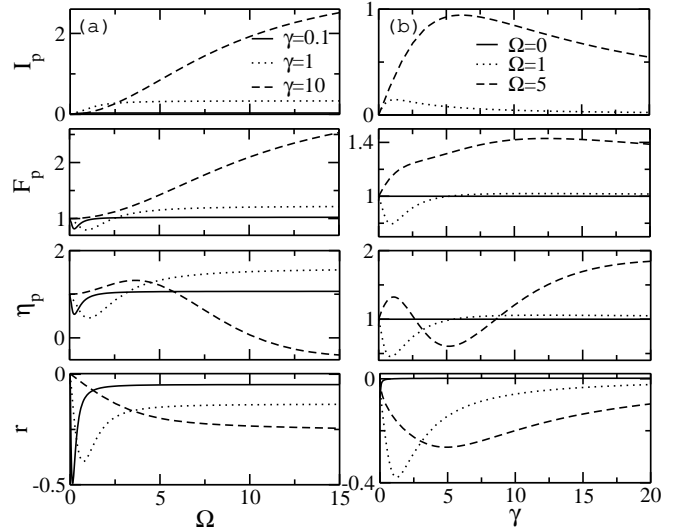}
\end{center}
\caption{\label{phudW}\small {\bf Level dependent tunneling:} Photonic current, Fano factor and skewness and electron-photon correlation coefficient as a function of (a) the field intensity, for different photon emission rates, and (b) the photon emission rate for different field intensities, for the case: $\Gamma_{1}=1$ and $\Gamma_{2}=10^{-5}$.
}
\end{figure}

In the absence of driving, the electrons tend to be transferred through the lower level (and the system is reduced to the single resonant level configuration\cite{hersfield}), so there is no chance for photon emission.
The introduction of the driving field populates the upper level thus reducing the electronic current, cf. Fig. \ref{eudW},  and giving a finite probability to photons to be emitted --thus acting as a {\it photon pump}, cf. Fig. \ref{phudW}a:
\begin{eqnarray}
I_e&=&\frac{\Gamma_{1}\left(\gamma^2+\gamma\Gamma_{1}+\Omega^2\right)}{2\gamma^2+2\gamma\Gamma_{1} +3 \Omega ^2}\\
I_p&=&\frac{\gamma\Omega^2}{2\gamma^2+2\gamma\Gamma_{1} +3\Omega ^2}.
\end{eqnarray}
The ac field modifies the electronic Fano factor typical from the single resonant level, $F_e=1/2$, without changing its sub-Poissonian character but for the range $\Gamma_{1}>\sqrt{2}\Omega\gg\gamma$:
\begin{equation}
F_e=1-\frac{2\left(\gamma^2(\gamma+\Gamma_{1})^2+\Omega^2\left(3\gamma^2-\Gamma_{1}^2\right)+2\Omega^4\right)}{\left(2 \gamma ^2+2 \Gamma _{1} \gamma +3 \Omega ^2\right)^2}.
\end{equation}
The photonic Fano factor:
\begin{equation}
F_p=1+\frac{2 \gamma  \Omega ^2 \left(\gamma ^2-5 \Gamma _{1} \gamma-2 \Gamma _{1}^2+\Omega ^2\right)}{\Gamma _{1} \left(2 \gamma ^2+2 \Gamma _{1} \gamma +3\Omega ^2\right)^2}
\end{equation}
can be turned from sub-Poissonian to super-Poissonian by increasing the field intensity if $\Gamma_{1}>\frac{1}{2}(\sqrt{33}+5)\gamma$. Otherwise, it will be always super-Poissonian. The electron-photon correlation, calculated from (\ref{c11G1_0}), is always negative.

The third electronic cumulant varies between $\eta_e=\frac{1}{4}$, for $\Omega=0$, and $\eta_e=\frac{7}{27}$ for the high intensity limit, but it shows a deep minimum for low voltages where it is negative, cf. Fig. \ref{eudW}. The photonic one is removed by the AC field from $\eta_p=1$ to the asymptotic limit:
\begin{equation}
\eta_p=1-\frac{2\gamma(\gamma-9\Gamma_{1})}{27\Gamma_{1}^2},
\end{equation}
for $\Omega\rightarrow\infty$. Then, the skewness of the photonic statistics can be tuned by the strength of the tunneling couplings.

%\begin{figure}[t]
%\begin{center}
%\includegraphics[width=3in,clip]{fephrGu000001.eps}
%\includegraphics[width=3in,clip]{etaephGu000001.eps}
%\end{center}
%\caption{\label{udW}\small {\bf Level dependent tunneling} (a) $F_e$ (solid), $F_p$ (dashed) and $r$ (dotted) and (b) $\eta_e$ (solid) and $\eta_p$ (dashed) as a function of the driving intensity in resonance for $\gamma=0.1$, $\Gamma_{2}=10^{-5}$ and $\Gamma_{1}=1$.
%}
%\end{figure}

\section{Conclusions}
\label{conclus}
A method for extracting the simultaneous counting statistics for electrons tunneling through an ac driven two level quantum dot and for photons emitted in the intradot electron relaxation processes is presented. It allows to calculate all the electronic and photonic cumulants as well as the correlation between fermionic and bosonic statistics, showing how they affect one to the other.
For instance, photon emission is shown to reduce the super-Poissonian electronic shot noise in the dynamical channel blockade regime. On the other hand, a purely quantum feature as is sub-Poissonian statistics in a two level photon source (resonance fluorescence) is lost as the electron is allowed to escape from the system. Our method can be applied to obtain the correlations between processes of different kind affecting to the same system as could be spin dependent transport or three terminal devices.

It is shown how the character of the electronic and photonic fluctuations can be manipulated by tuning the external parameters like the intensity of the AC field, the chemical potencial of the right lead or the tunneling barriers. By this kind of measurements, information about electron relaxation times can be obtained. All the combinations of sub- and super-Poissonian noise can be selected in this way\cite{PRL}.

We present an analysis of the electron-photon correlations which gives a more complete understanding of the dynamical behaviour of each concrete sample configuration and the importance of relaxation processes in transport properties. In this sense, a configuration with a maximal electron-photon correlation is proposed. Additionally, this configuration can serve as a probe for the photonic emission rate.

A triple quantum dot system is proposed in order to control tunneling through the central two level quantum dot, while the levels of the neightbour dots act as zero-dimensional leads. This way, assorted configurations which can be mapped to coherently or incoherently coupled double quantum dot systems, or quantum dots coupled to localized states can be achieved, providing a way to explore the effect of coherence in electronic transport.

\section*{Acknowledgments}
We thank M. B\"uttiker for comments. R.S. and G.P. were supported by the M.E.C. of Spain
through Grant No. MAT2005-00644.
T.B. acknowledges support by the German DFG project BR 1528/5-1 and the DAAD.

%%%%%%%%%%%%%%%%%%%%%%%%%%%%%%%%%%%%%%%%%%%%%%
%           APPENDIX
%%%%%%%%%%%%%%%%%%%%%%%%%%%%%%%%%%%%%%%%%%%%%%

\appendix

\begin{widetext}

\section{Photonic Resonance Fluorescence}
\label{aprf}
When the chemical potential is above the energies of both levels, so the probability for an electron in the upper level level to tunnel to the collector is $x\Gamma_{\rm R}$, the up to second order moments are given by the coefficients (considering the Taylor expansion around $x=0$):
\begin{eqnarray}
%c_{10}&=&\frac{\Omega ^2 \Gamma _{\rm L} \Gamma _{\rm R}}{\left(\gamma ^2+2 \Omega ^2\right) \left(\Gamma _{\rm L}+\Gamma _{\rm R}\right)}x+O\left(x^2\right)\\
c_{20}&=&-\frac{\Omega ^2 \Gamma _{\rm L}^2 \Gamma _{\rm R}^2 \left(\Omega ^2 \left(\gamma ^2+2 \Omega ^2\right)-\gamma  (\gamma -2 \Omega ) (\gamma +2 \Omega) \left(\Gamma _{\rm L}+\Gamma _{\rm R}\right)\right)}{2 \left(\left(\gamma ^2+2 \Omega ^2\right)^3 \left(\Gamma _{\rm L}+\Gamma_{\rm R}\right)^3\right)}x^2+O\left(x^3\right)\\
%c_{30}&=&\frac{\Omega ^2 \Gamma _{\rm L}^3 \Gamma _{\rm R}^3x^3}{4 \left(\gamma ^2+2 \Omega ^2\right)^5 \left(\Gamma _{\rm L}+\Gamma_{\rm R}\right)^5}\nonumber\\[-2.3mm]
%&&\\[-2.3mm]
%&&\times\left(2 \left(\gamma ^2+2 \Omega ^2\right)^2 \Omega ^4+\gamma  \left(\Gamma _{\rm L}+\Gamma _{\rm R}\right)\left(24 \Omega ^6+6 \gamma ^2 \Omega ^4-3 \gamma ^4\Omega ^2+\gamma  \left(\gamma ^4-10 \Omega ^2 \gamma ^2+48 \Omega^4\right)\left(\Gamma _{\rm L}+\Gamma _{\rm R}\right)\right)\right)+O\left(x^4\right)\nonumber
%c_{01}&=&\frac{\gamma  \Omega ^2}{\gamma ^2+2 \Omega ^2}-\frac{\gamma  \Omega ^2 \Gamma _{\rm R} \left(\Omega ^2+3 \gamma  \left(\Gamma _{\rm L}+\Gamma_{\rm R}\right)\right)}{2 \left(\left(\gamma ^2+2 \Omega ^2\right)^2 \left(\Gamma _{\rm L}+\Gamma _{\rm R}\right)\right)}x+O\left(x^2\right)\\
c_{02}&=&-\frac{3\gamma ^3 \Omega ^4}{\left(\gamma ^2+2 \Omega ^2\right)^3}
%\\[-2.5mm]
%&&\\[-2.5mm]
%&&
-\frac{\gamma ^2\Omega^4\Gamma_{\rm R}}{\left(\gamma ^2+2 \Omega ^2\right)^3}\left(\frac{\gamma(\gamma^2-16\Omega^2)}{4\left(\gamma ^2+2 \Omega ^2\right)(\Gamma_{\rm L}+\Gamma_{\rm R})}-\frac{\Omega^2}{4(\Gamma_{\rm L}+\Gamma_{\rm R})^2}-\frac{23\gamma^2-8\Omega^2}{2(\gamma ^2+2 \Omega ^2)}\right)x+O\left(x^2\right)\\
c_{11}&=&\frac{\gamma  \Omega ^2 \Gamma _{\rm L} \Gamma _{\rm R}
\left(\gamma  \left(\gamma ^2-10 \Omega ^2\right) \left(\Gamma
_{\rm L}+\Gamma _{\rm R}\right)-\Omega ^2\left(\gamma ^2+2 \Omega
^2\right)\right)}{2 \left(\gamma ^2+2 \Omega ^2\right)^3
\left(\Gamma _{\rm L}+\Gamma _{\rm R}\right)^2}x+O\left(x^2\right)
\end{eqnarray}
%\bea
%c_{20}&=&c_{10}\frac{ x \Gamma _{\rm L} \Gamma _{\rm R} \left(\gamma ^2+4 \Omega ^2+x \Gamma _{\rm R} \left(2 \gamma +x \Gamma _{\rm R}\right)\right)}{\left(\gamma +x\Gamma _{\rm R}\right) \left(\Gamma _{\rm L} \left(2 \left(\gamma ^2+2 \Omega ^2\right)+x \Gamma _{\rm R} \left(3 \gamma +x \Gamma _{\rm R}\right)\right)+\Gamma _{\rm R}\left(-(x-2) \gamma ^2-(x-4) \Omega ^2+x \Gamma _{\rm R} \left(x \Gamma _{\rm R}-(x-3) \gamma \right)\right)\right)}\\
%&&+c_{10}^2\frac{\left(-\gamma\left(\gamma ^2+2 \Omega ^2\right)-2 \Gamma _{\rm L} \left(5 \gamma ^2+4 \Omega ^2+x \Gamma _{\rm R} \left(8 \gamma +3 x \Gamma _{\rm R}\right)\right)-\Gamma_{\rm R} \left(-2 (x-5) \gamma ^2+8 \Omega ^2+x \Gamma _{\rm R} \left(-3 x \gamma +16 \gamma +6 x \Gamma _{\rm R}\right)\right)\right)}{\left(\gamma +x\Gamma _{\rm R}\right) \left(\Gamma _{\rm L} \left(2 \left(\gamma ^2+2 \Omega ^2\right)+x \Gamma _{\rm R} \left(3 \gamma +x \Gamma _{\rm R}\right)\right)+\Gamma _{\rm R}\left(-(x-2) \gamma ^2-(x-4) \Omega ^2+x \Gamma _{\rm R} \left(x \Gamma _{\rm R}-(x-3) \gamma \right)\right)\right)}
%\eea
%\end{widetext}

%\subsection{Undriven case}
For the undriven and high AC intensity limits, one can give short expressions without having to do the Taylor expansion around $x=0$. In the undriven case, $\Omega=0$, the electron-photon correlation is given by: 
\be
c_{11}=\frac{c_{01}\left(x \Gamma _{\rm L} \Gamma _{\rm R}-2 c_{10} \left(\gamma +2 \Gamma _{\rm L}+2 \Gamma _{\rm R}\right)\right)}{\Gamma _{\rm L} \left(2 \gamma +x \Gamma _{\rm R}\right)+\Gamma_{\rm R} \left(x \Gamma _{\rm R}-(x-2) \gamma \right)}-\frac{8 c_{10}c_{01}}{\gamma +x\Gamma _{\rm R}}.
\ee
%\subsection{High Intensity Limit}
In the opposite case, $\Omega\rightarrow\infty$, on obtains the electron-photon correlation from the coefficient:
\be
c_{11}=\frac{4 x c_p \Gamma _{\rm L} \Gamma _{\rm R}+c_e \left(\left(\gamma -4 c_p\right) \left(\gamma +4 \Gamma _{\rm L}\right)-\left((x-4) \gamma +16 c_p\right)\Gamma _{\rm R}\right)}{\left(4 \Gamma _{\rm L}-(x-4) \Gamma _{\rm R}\right) \left(\gamma +x \Gamma _{\rm R}\right)}
\ee
and the skewness of the electronic and photonic statistics from:
\begin{eqnarray}
%c_{20}&=&-\frac{8 x^2 \Gamma_{\rm L}^2 \Gamma _{\rm R}^2}{\left(4 \Gamma _{\rm L}+(4-x) \Gamma _{\rm R}\right)^3}\\
%c_{02}&=&\frac{x \gamma ^2 \Gamma _{\rm R} \left(2 \Gamma _{\rm L}+(2-x) \Gamma _{\rm R}\right)}{\left(4 \Gamma _{\rm L}-(x-4) \Gamma _{\rm R}\right)^3}\\
c_{30}&=&\frac{64 x^3 \Gamma _{\rm L}^3 \Gamma _{\rm R}^3}{\left(4 \Gamma _{\rm L}+(4-x) \Gamma _{\rm R}\right)^5}\\
c_{03}&=&-\frac{x \gamma ^3 \Gamma _{\rm R} \left(4 \Gamma _{\rm L}+(4-3 x) \Gamma _{\rm R}\right) \left(2 \Gamma _{\rm L}+(2-x)\Gamma _{\rm R}\right)}{\left(4 \Gamma _{\rm L}+(4-x)\Gamma _{\rm R}\right)^5}
\end{eqnarray}

%\begin{widetext}
\section{Dynamical Channel Blockade regime.}
\label{apdcb}
If the chemical potential of the collector is between the energies of each level, so the occupation of the lower level avoids electrons from tunneling through the upper one, until it is extracted with a rate $x\Gamma_{\rm R}$, the second order correlations are given by (considering $x=0$):
\bea
c_{20}&=&c_{10} \frac{\Gamma _{\rm L} \Gamma _{\rm R} \left(\gamma ^2+2 \Gamma _{\rm R} \gamma +4 \Omega ^2+\Gamma _{\rm R}^2\right)}{\left(\gamma +\Gamma _{\rm R}\right) \left(\Gamma _{\rm R} \left(\gamma ^2+2 \Gamma _{\rm R} \gamma +3 \Omega^2+\Gamma _{\rm R}^2\right)+\Gamma _{\rm L} \left(2 \gamma ^2+3 \Gamma _{\rm R} \gamma +4 \Omega ^2+\Gamma _{\rm R}^2\right)\right)}\nonumber\\[-2.5mm]
&&\\[-2.5mm]
&&-c_{10}^2\frac{\gamma ^3+2 \Omega ^2 \gamma+13 \Gamma _{\rm R}^2 \gamma +6 \Gamma _{\rm R}^3+8 \left(\gamma ^2+\Omega ^2\right) \Gamma _{\rm R}+2 \Gamma _{\rm L} \left(5 \gamma ^2+8 \Gamma _{\rm R} \gamma +4\Omega ^2+3 \Gamma _{\rm R}^2\right)}{\left(\gamma +\Gamma _{\rm R}\right) \left(\Gamma _{\rm R} \left(\gamma ^2+2 \Gamma _{\rm R} \gamma +3 \Omega^2+\Gamma _{\rm R}^2\right)+\Gamma _{\rm L} \left(2 \gamma ^2+3 \Gamma _{\rm R} \gamma +4 \Omega ^2+\Gamma _{\rm R}^2\right)\right)}\nonumber\\
c_{02}&=&c_{01}\frac{\gamma  \Omega ^2 \left(\gamma +4 \Gamma _{\rm L}+3 \Gamma_{\rm R}\right)-c_{01} \left(\gamma ^3+2 \Omega ^2 \gamma +13 \Gamma _{\rm R}^2 \gamma +6 \Gamma _{\rm R}^3+8 \left(\gamma ^2+\Omega ^2\right) \Gamma _{\rm R}+2\Gamma _{\rm L} \left(5 \gamma ^2+8 \Gamma _{\rm R} \gamma +4 \Omega ^2+3 \Gamma _{\rm R}^2\right)\right)}{\left(\gamma +\Gamma _{\rm R}\right) \left(\Gamma _{\rm R} \left(\gamma ^2+2 \Gamma _{\rm R} \gamma +3 \Omega ^2+\Gamma _{\rm R}^2\right)+\Gamma_{\rm L} \left(2 \gamma ^2+3 \Gamma _{\rm R} \gamma +4 \Omega ^2+\Gamma _{\rm R}^2\right)\right)}\\
c_{11}&=&\frac{-\gamma  \Omega ^2 \left(\gamma +4 \Gamma _{\rm L}+3 \Gamma _{\rm R}\right)+c_{01} \Gamma _{\rm L} \Gamma _{\rm R} \left(\gamma ^2+2 \Gamma _{\rm R} \gamma +4 \Omega ^2+\Gamma _{\rm R}^2\right)}{\left(\gamma+\Gamma _{\rm R}\right) \left(\Gamma _{\rm R} \left(\gamma ^2+2 \Gamma _{\rm R} \gamma +3 \Omega ^2+\Gamma _{\rm R}^2\right)+\Gamma _{\rm L} \left(2 \gamma ^2+3 \Gamma _{\rm R}\gamma +4 \Omega ^2+\Gamma _{\rm R}^2\right)\right)}\\
&&-c_{10}\frac{2 c_{01} \left(\gamma ^3+2 \Omega ^2\gamma +6 \Gamma _{\rm R}^3+\left(13 \gamma +6 \Gamma _{\rm L}\right) \Gamma _{\rm R}^2+2 \left(5 \gamma ^2+4 \Omega ^2\right) \Gamma _{\rm L}+8 \left(\gamma ^2+2\Gamma _{\rm L} \gamma +\Omega ^2\right) \Gamma _{\rm R}\right)}{\left(\gamma+\Gamma _{\rm R}\right) \left(\Gamma _{\rm R} \left(\gamma ^2+2 \Gamma _{\rm R} \gamma +3 \Omega ^2+\Gamma _{\rm R}^2\right)+\Gamma _{\rm L} \left(2 \gamma ^2+3 \Gamma _{\rm R}\gamma +4 \Omega ^2+\Gamma _{\rm R}^2\right)\right)}
\eea
%and
%\begin{eqnarray}
%c_{30}&=&\frac{2 \Gamma\Omega ^2}{\left(7\Omega ^2+(\gamma +\Gamma ) (3 \gamma +2 \Gamma )\right)^5}\left(\Gamma ^2 (\gamma +\Gamma )^4 (3 \gamma +2 \Gamma )^2-2 \Gamma  (\gamma +\Gamma )^2 (3 \gamma +2 \Gamma )\left(3 \gamma ^2+20 \Gamma  \gamma +13 \Gamma ^2\right) \Omega ^2\right.\\
%&&\left.+\left(8 \gamma ^4+62 \Gamma\gamma ^3+545 \Gamma ^2 \gamma ^2+700 \Gamma ^3 \gamma +241 \Gamma ^4\right) \Omega ^4+4 \left(8 \gamma ^2+46 \Gamma  \gamma +29 \Gamma ^2\right) \Omega ^6+32 \Omega ^8\right)\nonumber\\
%c_{03}&=&\frac{3 \gamma ^3 \Omega ^6 \left(-5 \Omega ^4-\left(3\gamma ^2+110 \Gamma  \gamma +503 \Gamma ^2\right) \Omega^2+\Gamma  \left(3 \gamma^3+1396 \Gamma  \gamma ^2+2171 \Gamma ^2\gamma +850 \Gamma ^3\right)\right)}{\Gamma ^2 \left(7 \Omega^2+(\gamma +\Gamma ) (3 \gamma +2 \Gamma )\right)^5}
%\end{eqnarray}
%\subsection{Undriven case.}

For small $x$, in the undriven configuration, $\Omega=0$, the electron-photon correlation is given by: 
\begin{equation}
c_{11}=\frac{\gamma  \Gamma _{\rm L} \Gamma _{\rm R} \left(\gamma +\Gamma _{\rm R}\right) \left(2 \Gamma _{\rm L}+\Gamma _{\rm R}\right) x}{\left(\Gamma _{\rm R} \left(\gamma +\Gamma_{\rm R}\right)+\Gamma _{\rm L} \left(2 \gamma +\Gamma _{\rm R}\right)\right)^2}+O\left(x^2\right).%\stackrel{\Gamma_{\rm L}=\Gamma_{\rm R}}{\longrightarrow}3\gamma\Gamma x\frac{\gamma+\Gamma}{(3\gamma+2\Gamma)^2}
\end{equation}
%\bea
%c_{11}=\frac{\Gamma _{\rm L} \Gamma _{\rm R} \left(x \gamma  \left(\gamma +(x+1) \Gamma _{\rm R}\right)+c_p \left(9 x \gamma +\gamma +(x (x+10)+1) \Gamma _{\rm R}\right)\right)-2 c_e \left(c_p \left(\gamma ^2+(x+7) \Gamma _{\rm R} \gamma +2 (x+3) \Gamma _{\rm R}^2+2 \Gamma _{\rm L} \left(5 \gamma +3 (x+1) \Gamma_{\rm R}\right)\right)-2 x \gamma  \Gamma _{\rm L} \Gamma _{\rm R}\right)}{\left(\gamma +(x+1) \Gamma _{\rm R}\right) \left(\Gamma _{\rm R} \left(\gamma +\Gamma_{\rm R}\right)+\Gamma _{\rm L} \left(2 \gamma +(x+1) \Gamma _{\rm R}\right)\right)}
%\eea
%\subsection{High intensity limit}
In the opposite limit, $\Omega\rightarrow\infty$, we find short expressions for the electron-photon correlation: 
\begin{eqnarray}
%c_{20}&=&-\frac{(x+1)^2\Gamma _{\rm L}^3 \Gamma _{\rm R}^3}{\left(4 \Gamma _{\rm L}+(3-x) \Gamma _{\rm R}\right)^3}\\
%c_{02}&=&\frac{(x+1) \gamma^2 \Gamma_{\rm R}\left(2\Gamma_{\rm L}+(1-x)\Gamma _{\rm R}\right)}{\left(4 \Gamma_{\rm L}+(3-x)\Gamma _{\rm R}\right)^3}\\
c_{11}&=&-\frac{2(x+1)\gamma\Gamma _{\rm L}\Gamma _{\rm R}(4\Gamma _{\rm L}+(1-3x)\Gamma_{\rm R})}{\left(4 \Gamma _{\rm L}+(3-x) \Gamma _{\rm R}\right)^3},
\end{eqnarray}
and the third order moments:
\begin{eqnarray}
c_{30}&=&\frac{64 (x+1)^3 \Gamma _{\rm L}^3 \Gamma _{\rm
R}^3}{\left(4 \Gamma _{\rm L}+(3-x) \Gamma _{\rm R}\right)^5}\\
c_{03}&=&-\frac{(x+1) \gamma ^3 \Gamma _{\rm R} \left(8 \Gamma
_{\rm L}^2+2 (3-5 x) \Gamma _{\rm R} \Gamma _{\rm L}+\left(3 x^2-4
x+1\right) \Gamma _{\rm R}^2\right)}{\left(4 \Gamma_{\rm L}+(3-x)
\Gamma _{\rm R}\right)^5}.
\end{eqnarray}
%\end{widetext}

%\begin{widetext}
\section{Both levels in the transport window}
\label{aphb}
If the chemical potential of the collector is below the energy of both levels, the photonic shot noise and the electron-photon correlation can be obtained from: 
\bea
c_{02}&=&c_{01}\frac{ \gamma  \left(\left(\gamma +2 \Gamma _{\rm R}\right) \Omega ^2+4 \Gamma _{\rm L} \left(\Omega ^2+\Gamma _{\rm R} \left(\gamma +2 \Gamma_{\rm R}\right)\right)\right)}{\left(2 \Gamma_{\rm L}+\Gamma _{\rm R}\right) \left(\gamma +2 \Gamma _{\rm R}\right) \left(\gamma ^2+3 \Gamma _{\rm R} \gamma +2 \Omega ^2+2 \Gamma _{\rm R}^2\right)}\nonumber\\[-2.5mm]
&&\\[-2.5mm]
&&-c_{01}^2\frac{\gamma ^3+2 \Omega ^2 \gamma +2 \Gamma _{\rm R} \left(5 \gamma ^2+12 \Gamma _{\rm R} \gamma +4 \Omega ^2+8 \Gamma_{\rm R}^2\right)+2 \Gamma _{\rm L} \left(5 \gamma ^2+4 \Omega ^2+4 \Gamma _{\rm R} \left(4 \gamma +3 \Gamma _{\rm R}\right)\right)}{\left(2 \Gamma_{\rm L}+\Gamma _{\rm R}\right) \left(\gamma +2 \Gamma _{\rm R}\right) \left(\gamma ^2+3 \Gamma _{\rm R} \gamma +2 \Omega ^2+2 \Gamma _{\rm R}^2\right)}\nonumber\\
\label{hbeph}
c_{11}&=&\frac{1}{\left(2 \Gamma _{\rm L}+\Gamma _{\rm R}\right) \left(\gamma +2 \Gamma _{\rm R}\right) \left(\gamma ^2+3 \Gamma _{\rm R} \gamma +2\Omega ^2+2 \Gamma _{\rm R}^2\right)}\left[\Gamma _{\rm L} \Gamma _{\rm R}\gamma  \left(\gamma +2 \Gamma _{\rm R}\right)^2\right.\nonumber\\
&&\left.+2c_{01}\Gamma _{\rm L} \Gamma _{\rm R}\left(5 \gamma ^2+4 \Omega ^2+4 \Gamma _{\rm R} \left(4\gamma+3\Gamma _{\rm R}\right)\right)\right.\nonumber\\[-2.5mm]
\\[-2.5mm]
&&\left.+c_{10}\gamma  \left(\left(\gamma +2 \Gamma _{\rm R}\right) \Omega ^2+4 \Gamma _{\rm L} \left(\Omega ^2+\Gamma _{\rm R}\left(\gamma +2 \Gamma _{\rm R}\right)\right)\right)\right.\nonumber\\
&&\left.-2c_{10} c_{01}\left(\gamma ^3+2 \Omega ^2 \gamma +2 \Gamma _{\rm R} \left(5 \gamma ^2+12 \Gamma _{\rm R} \gamma+4 \Omega ^2+8 \Gamma _{\rm R}^2\right)+2 \Gamma _{\rm L} \left(5 \gamma ^2+4 \Omega ^2+4 \Gamma _{\rm R} \left(4 \gamma +3 \Gamma_{\rm R}\right)\right)\right)\right]\nonumber
\eea
%\begin{equation}
%\label{hbeph}
%c_{11}=-\gamma\frac{\Gamma(\Gamma-5\gamma)(\gamma+2\Gamma)^2+2 \left(\gamma ^2+16 \Gamma  \gamma +4 \Gamma ^2\right) \Omega ^2+4 \Omega ^4}{27 \left(2 \Omega ^2+(\gamma +\Gamma ) (\gamma +2 \Gamma )\right)^2}.
%\end{equation}
We also obtain the skewness of the photonic statistics for the undriven case, $\Omega=0$: 
%\subsection{Undriven Case}
%\label{aphbuc}
\begin{eqnarray}
%c_{01}&=&\frac{\gamma  \Gamma _{\rm L} \Gamma _{\rm R}}{\left(\gamma +\Gamma _{\rm R}\right) \left(2 \Gamma _{\rm L}+\Gamma _{\rm R}\right)} \stackrel{\Gamma_{\rm L}=\Gamma_{\rm R}}{\longrightarrow}\frac{\gamma\Gamma}{3(\gamma+\Gamma)}\\
%c_{02}&=&-\frac{\gamma ^2 \Gamma _{\rm L}^2 \Gamma _{\rm R}^2\left(\gamma +2 \Gamma _{\rm L}+2 \Gamma _{\rm R}\right)}{\left(\gamma +\Gamma _{\rm R}\right)^3\left(2\Gamma_{\rm L}+\Gamma _{\rm R}\right)^3}\\
c_{03}&=&\frac{\gamma ^3 \Gamma _{\rm L}^3 \Gamma _{\rm R}^3
\left(2 \gamma ^2+8 \Gamma _{\rm L}^2+7 \Gamma _{\rm R}
\left(\gamma +\Gamma _{\rm R}\right)+2 \Gamma _{\rm L} \left(3
\gamma
   +7 \Gamma _{\rm R}\right)\right)}{\left(\gamma +\Gamma _{\rm R}\right)^5 \left(2 \Gamma _{\rm L}+\Gamma _{\rm R}\right)^5},
\end{eqnarray}
%\subsection{High intensity limit}
and in the high AC intensity regime:
\begin{eqnarray}
%c_{02}&=&\frac{\gamma ^2 \Gamma _{\rm L} \Gamma _{\rm R}}{2 \left(2 \Gamma _{\rm L}+\Gamma _{\rm R}\right)^3}\\
c_{03}&=&\frac{\gamma ^3 \Gamma _{\rm L} \Gamma _{\rm R} \left(\Gamma _{\rm R}-2 \Gamma _{\rm L}\right)}{4 \left(2 \Gamma _{\rm L}+\Gamma _{\rm R}\right)^5}
\end{eqnarray}
%\end{widetext}

%\begin{widetext}
\section{Selective tunneling configuration}
\label{apstep}
In the configuration describen in Sec. \ref{secselect}, the electronic and photonic correlations are given by:
\begin{eqnarray}\label{stepces}
%c_{10}&=&\frac{\Gamma _{\rm L} \Gamma _{\rm R} \left(\gamma^2+\Gamma _{\rm R} \gamma +\Omega ^2\right)}{\Gamma _{\rm R}\left(\gamma ^2+\Gamma _{\rm R} \gamma +\Omega^2\right)+\Gamma _{\rm L} \left(\gamma ^2+2 \Gamma _{\rm R} \gamma +2 \Omega ^2+\Gamma _{\rm R}^2\right)}\\
%\frac{\Gamma  \left(\gamma ^2+\Gamma  \gamma +\Omega ^2\right)}{2 \gamma ^2+3 \Gamma  \gamma +\Gamma ^2+3 \Omega ^2}\\
%c_{01}&=&\frac{\gamma  \Gamma _{\rm L} \left(\Omega ^2+\Gamma_{\rm R} \left(\gamma +\Gamma _{\rm R}\right)\right)}{\Gamma _{\rm R} \left(\gamma ^2+\Gamma _{\rm R} \gamma +\Omega^2\right)+\Gamma _{\rm L} \left(\gamma ^2+2 \Gamma _{\rm R} \gamma +2 \Omega ^2+\Gamma _{\rm R}^2\right)}\\
%\frac{\gamma  \left(\Gamma ^2+\gamma  \Gamma +\Omega ^2\right)}{2 \gamma ^2+3 \Gamma  \gamma +\Gamma ^2+3 \Omega ^2}\\
c_{20}&=&\frac{c_{10}2 \Gamma _{\rm L} \Gamma _{\rm R} \left(2 \gamma ^2+2 \Gamma _{\rm R} \gamma +\Omega ^2\right)}{\left(\gamma +\Gamma _{\rm R}\right) \left(\Gamma _{\rm R} \left(\gamma ^2+\Gamma _{\rm R} \gamma +\Omega^2\right)+\Gamma _{\rm L} \left(\gamma ^2+2 \Gamma _{\rm R} \gamma +2 \Omega ^2+\Gamma _{\rm R}^2\right)\right)}\nonumber\\[-2.5mm]
&&\\[-2.5mm]
&&-c_{10}^2 \frac{\gamma ^3+2 \Omega ^2 \gamma +\Gamma _{\rm R}\left(7 \gamma ^2+7 \Gamma _{\rm R} \gamma +4 \Omega ^2+\Gamma _{\rm R}^2\right)+\Gamma _{\rm L} \left(5 \gamma ^2+4 \Omega ^2+5 \Gamma _{\rm R} \left(2 \gamma+\Gamma _{\rm R}\right)\right)}{\left(\gamma +\Gamma _{\rm R}\right) \left(\Gamma _{\rm R} \left(\gamma ^2+\Gamma _{\rm R} \gamma +\Omega^2\right)+\Gamma _{\rm L} \left(\gamma ^2+2 \Gamma _{\rm R} \gamma +2 \Omega ^2+\Gamma _{\rm R}^2\right)\right)}\nonumber\\
c_{02}&=&\frac{c_{01}\gamma  \left(\left(\gamma +\Gamma _{\rm
R}\right) \Omega ^2+2 \Gamma _{\rm L} \left(\Omega ^2+2 \Gamma
_{\rm R} \left(\gamma +\Gamma
   _{\rm R}\right)\right)\right)}{\left(\gamma +\Gamma_{\rm R}\right) \left(\Gamma _{\rm R} \left(\gamma ^2+\Gamma _{\rm R} \gamma +\Omega ^2\right)+\Gamma _{\rm L} \left(\gamma ^2+2 \Gamma _{\rm R} \gamma +2 \Omega ^2+\Gamma_{\rm R}^2\right)\right)}\nonumber\\[-2.5mm]
&&\\[-2.5mm]
&&-c_{01}^2\frac{\gamma ^3+2 \Omega ^2 \gamma +\Gamma _{\rm R}
\left(7 \gamma ^2+7 \Gamma _{\rm R} \gamma +4 \Omega ^2+\Gamma
   _{\rm R}^2\right)+\Gamma _{\rm L} \left(5 \gamma ^2+4 \Omega ^2+5 \Gamma _{\rm R} \left(2 \gamma +\Gamma _{\rm R}\right)\right)}{\left(\gamma +\Gamma_{\rm R}\right) \left(\Gamma _{\rm R} \left(\gamma ^2+\Gamma _{\rm R} \gamma +\Omega ^2\right)+\Gamma _{\rm L} \left(\gamma ^2+2 \Gamma _{\rm R} \gamma +2 \Omega ^2+\Gamma_{\rm R}^2\right)\right)}\nonumber
%c_{20}&=&-\frac{\Gamma  \left(\gamma ^2+\Gamma  \gamma +\Omega ^2\right) \left(\gamma ^4+4 \Gamma  \gamma ^3+5 \Gamma ^2 \gamma ^2+3 \Omega ^2 \gamma^2+2 \Gamma ^3 \gamma +3 \Gamma  \Omega ^2 \gamma +2 \Omega ^4+4 \Gamma ^2 \Omega ^2\right)}{\left(2 \gamma ^2+3 \Gamma  \gamma +\Gamma ^2+3\Omega ^2\right)^3}\\
%c_{02}&=&-\frac{\gamma ^2 \left(\Gamma ^2+\gamma  \Gamma +\Omega ^2\right) \left(2 \Gamma ^4-\Omega ^2 \Gamma ^2+\gamma ^3 \Gamma -\Omega ^4+\gamma ^2\left(4 \Gamma ^2-\Omega ^2\right)+\gamma  \left(5 \Gamma ^3+6 \Omega ^2 \Gamma \right)\right)}{\Gamma  \left(2 \gamma ^2+3 \Gamma  \gamma+\Gamma ^2+3 \Omega ^2\right)^3}\\
\end{eqnarray}
and, considering $\Gamma_{\rm L}=\Gamma_{\rm R}=\Gamma$, for
simplicity:
\begin{equation}
c_{11}
=\frac{\gamma  \left(\Gamma  (\gamma +\Gamma )^3 \left(2 \gamma
^2+\Gamma ^2\right)-\gamma  (\gamma -11 \Gamma ) \Gamma
(\gamma+\Gamma ) \Omega ^2-\left(\gamma ^2+\Gamma  \gamma -4
\Gamma ^2\right) \Omega ^4-\Omega ^6\right)}{\left(3 \Omega
^2+(\gamma +\Gamma ) (2 \gamma+\Gamma )\right)^3}.
\end{equation}
%\end{widetext}

%\subsection{Undriven case}
In the undriven case, $\Omega=0$, the third order coefficents are:
\begin{eqnarray}
%c_{10}&=&c_{01}=\frac{\gamma\Gamma_{\rm L}\Gamma_{\rm R}}{\gamma\Gamma_{\rm L}+\gamma\Gamma_{\rm R}+\Gamma_{\rm L}\Gamma_{\rm R}}\\
%c_{20}&=&c_{02}=-\frac{\gamma^2\Gamma_{\rm L}^2\Gamma_{\rm R}^2(\gamma+\Gamma_{\rm L}+\Gamma_{\rm R})}{(\gamma\Gamma_{\rm L}+\gamma\Gamma_{\rm R}+\Gamma_{\rm L}\Gamma_{\rm R})^3}\\
%c_{11}&=&\gamma\Gamma_{\rm L}\Gamma_{\rm R}\frac{\gamma^2\Gamma_{\rm L}^2+\gamma^2\Gamma_{\rm R}^2+\Gamma_{\rm L}^2\Gamma_{\rm R}^2}{(\gamma\Gamma_{\rm L}+\gamma\Gamma_{\rm R}+\Gamma_{\rm L}\Gamma_{\rm R})^3}\\
c_{30}&=&c_{03}=\gamma ^3 \Gamma _{\rm L}^3 \Gamma _{\rm R}^3
\frac{2 \left(\gamma ^2+\Gamma _{\rm L}^2+\Gamma _{\rm
R}^2\right)+3(\gamma\Gamma_{\rm L}+\gamma\Gamma_{\rm
R}+\Gamma_{\rm L}\Gamma_{\rm R})}{(\gamma\Gamma_{\rm
L}+\gamma\Gamma_{\rm R}+\Gamma_{\rm L}\Gamma_{\rm R})^5}
\end{eqnarray}

%\subsection{High intensity limit}
%\label{apstephb}
%\begin{eqnarray}
%c_{20}&=&-\frac{2 \Gamma _{\rm L}^2 \Gamma _{\rm R}^2}{\left(2\Gamma _{\rm L}+\Gamma _{\rm R}\right)^3}\\
%c_{02}&=&\frac{\gamma^2 \Gamma _{\rm L} \Gamma _{\rm R}}{\left(2 \Gamma _{\rm L}+\Gamma _{\rm R}\right)^3}\\
%c_{11}&=&\frac{\gamma^2 \Gamma _{\rm L} \Gamma _{\rm R}\left(\Gamma _{\rm R}-2 \Gamma_{\rm L}\right)}{\left(2 \Gamma _{\rm L}+\Gamma _{\rm R}\right)^3}
%\end{eqnarray}
%\begin{eqnarray}
%c_{30}&=&\frac{8 \Gamma _{\rm L}^3 \Gamma _{\rm R}^3}{\left(2\Gamma _{\rm L}+\Gamma _{\rm R}\right)^5}\\
%c_{03}&=&\frac{\gamma^3 \Gamma _{\rm L} \Gamma _{\rm R} \left(\Gamma _{\rm R}-2 \Gamma_{\rm L}\right)}{\left(2 \Gamma _{\rm L}+\Gamma _{\rm R}\right)^5}
%\end{eqnarray}

%\begin{widetext}
\section{Level dependent tunneling}
\label{apseltun}

%\subsection{Undriven case}
%Second order coefficients:
The tunneling between the leads and the quantum dot may depend in the involved level of the quantum dot. We consider here the case $\Gamma_{i{\rm L}}=\Gamma_{i{\rm R}}=\Gamma_{i}$ , where $i=\{1,2\}$. 
In the undriven case ($\Omega=0$), the electron-photon correlation is determined by the coefficient:
\bea
%c_{20}&=&\frac{\Gamma_1 \left(\gamma +\Gamma _2\right) \left(\Gamma _1+\Gamma _2\right) \left(\Gamma _2 \Gamma _1^3-\left(\gamma +2 \Gamma _2\right)^2\Gamma _1^2+\Gamma _2 \left(-\gamma ^2-\Gamma _2 \gamma +\Gamma _2^2\right) \Gamma _1+\gamma  \Gamma _2^3\right)}{\left(2 \gamma  \Gamma _1+\left(\gamma +3 \Gamma _1\right) \Gamma _2\right)^3}\\
%c_{02}&=&-\frac{\gamma ^2 \Gamma _1^2 \Gamma _2^2 \left(\gamma +2 \Gamma _1+2 \Gamma _2\right)}{\left(2 \gamma  \Gamma _1+\left(\gamma +3 \Gamma_1\right) \Gamma _2\right)^3}\\
c_{11}&=&\frac{\gamma  \Gamma _1 \Gamma _2 \left(\left(\gamma -\Gamma _1\right) \Gamma _2^3+\left(\gamma +\Gamma _1\right)^2 \Gamma _2^2+\Gamma _1\left(2 \gamma ^2+3 \Gamma _1 \gamma -\Gamma _1^2\right) \Gamma _2+2 \gamma  \left(\gamma -\Gamma _1\right) \Gamma _1^2\right)}{\left(2\gamma  \Gamma _1+\left(\gamma +3 \Gamma _1\right) \Gamma _2\right)^3}.
\eea
%\subsection{$\Gamma_1\ll\Gamma_2$}
%\label{G1_0}
We can give general expressions for the electron-photon correlation when tunneling through the one of the levels is supressed. For $\Gamma_1\ll\Gamma_2$:
%Second order coefficients:
\bea
%c_{20}&=&-\frac{\Omega ^2 \Gamma _2 \left(\Omega ^2 \left(\gamma ^2+2 \Omega ^2\right)-\Gamma _2 \left(\gamma ^3-2 \Omega ^2 \gamma +\Gamma _2 \left(2\gamma ^2+\Gamma _2 \gamma +\Omega ^2\right)\right)\right)}{\left(\gamma ^2+\Gamma _2 \gamma +3 \Omega ^2\right)^3}\\
%c_{02}&=&-\frac{\gamma ^2 \Omega ^4 \left(2 \Gamma _2 \left(2 \gamma +\Gamma _2\right)-\Omega ^2\right)}{\Gamma _2 \left(\gamma ^2+\Gamma _2 \gamma +3\Omega ^2\right)^3}\\
\label{c11G1_0}
c_{11}&=&\gamma\Omega^2\frac{\gamma\Gamma_{2}(\gamma+\Gamma_{2})^2-\Omega^2(\gamma^2+\Gamma_{2}^2+6\gamma\Gamma_{2})-\Omega^4}{(\gamma^2+3\Omega^2+\gamma\Gamma_{2})^3}
\eea
%The third ones:
%\bea
%c_{30}&=&\frac{\Omega ^2 \Gamma _2}{\left(\gamma ^2+\Gamma _2 \gamma +3 \Omega ^2\right)^5}\left(2 \left(\gamma ^2+2 \Omega ^2\right)^2 \Omega ^4+\Gamma _2 \left(13 \gamma  \Omega ^6+2 \gamma ^3 \Omega ^4-3\gamma ^5 \Omega ^2+\right.\right.\nonumber\\
%&&\left.\left.\Gamma _2 \left(\gamma ^6-13 \Omega ^2 \gamma ^4+5 \Omega ^4 \gamma ^2+\Omega ^6+\Gamma _2 \left(4 \gamma ^5-17 \Omega^2 \gamma ^3-4 \Omega ^4 \gamma +\Gamma _2 \left(6 \gamma ^4-7 \Omega ^2 \gamma ^2+\Gamma _2 \left(4 \gamma +\Gamma _2\right) \gamma^2-\Omega ^4\right)\right)\right)\right)\right)\nonumber\\
%c_{03}&=&\frac{\gamma ^3 \Omega ^6 \left(\Gamma _2 \left(\gamma ^3-14 \Omega ^2 \gamma +\Gamma _2 \left(30 \gamma ^2+29 \Gamma _2 \gamma -17 \Omega ^2+8\Gamma _2^2\right)\right)-\Omega ^2 \left(\gamma ^2+\Omega ^2\right)\right)}{\Gamma_2^2 \left(\gamma ^2+\Gamma _2 \gamma +3 \Omega^2\right)^5}.
%\eea
%\subsection{$\Gamma_2\ll\Gamma_1$}
%\label{G2_0}
%Second order coefficients:
and, for $\Gamma_2\ll\Gamma_1$:
\bea
%c_{20}&=&-\frac{\Gamma _1 \left(\gamma ^2+\Gamma _1 \gamma +\Omega ^2\right) \left(\gamma ^4+2 \Gamma _1 \gamma ^3+3 \Omega ^2 \gamma ^2+2 \Omega^4+(\gamma -\Omega ) (\gamma +\Omega ) \Gamma _1^2\right)}{\left(2 \gamma ^2+2 \Gamma _1 \gamma +3 \Omega ^2\right)^3}\\
%c_{02}&=&\frac{\gamma ^2 \Omega ^4 \left(\gamma ^2-5 \Gamma _1 \gamma +\Omega ^2-2 \Gamma _1^2\right)}{\Gamma _1 \left(2 \gamma ^2+2 \Gamma _1 \gamma +3\Omega ^2\right)^3}\\
\label{c11G2_0}
c_{11}&=&-\frac{\gamma  \Omega ^2 \left(2 \gamma  \Gamma _{1}^3+\left(8 \gamma ^2+\Omega ^2\right) \Gamma _{1}^2+\left(6 \gamma ^3+4 \Omega ^2 \gamma
   \right) \Gamma _{1}+\Omega ^2 \left(\gamma ^2+\Omega ^2\right)\right)}{\left(2 \gamma ^2+2 \Gamma _{1} \gamma +3 \Omega ^2\right)^3}.
\eea
%The third order ones:
%\bea
%c_{30}&=&\frac{\Gamma _1}{\left(2 \gamma ^2+2 \Gamma_1 \gamma +3 \Omega ^2\right)^5}\left(2 \left(\gamma ^2+2 \Omega ^2\right)^2 \left(\gamma ^2+\Omega ^2\right)^3+\right.\nonumber\\
%&&\left.\Gamma _1 \left(\gamma  \left(10 \gamma ^4+30\Omega ^2 \gamma ^2+17 \Omega ^4\right) \left(\gamma ^2+\Omega ^2\right)^2+\Gamma _1 \left(\left(\gamma ^2+\Omega ^2\right) \left(20 \gamma^6+30 \Omega ^2 \gamma ^4+8 \Omega ^4 \gamma ^2+\Omega ^6\right)+\right.\right.\right.\nonumber\\[-2.5mm]
%&&\\[-2.5mm]
%&&\left.\left.\left.\Gamma _1 \left(20 \gamma ^7-4 \Omega ^2 \gamma ^5-24 \Omega ^4 \gamma ^3-2\Omega ^6 \gamma -\Gamma _1 \left(-10 \gamma ^6+30 \Omega ^2 \gamma ^4+16 \Omega ^4 \gamma ^2+\right.\right.\right.\right.\right.\nonumber\\
%&&\left.\left.\left.\left.\left.\Gamma _1 \left(-2 \gamma ^4+14 \Omega ^2\gamma ^2+2 \Omega ^2 \Gamma _1 \gamma +3 \Omega ^4\right) \gamma +\Omega ^6\right)\right)\right)\right)\right)\nonumber\\
%c_{03}&=&-\frac{\gamma ^3 \Omega ^6 \left(\left(\gamma ^2+\Omega ^2\right) \Omega ^2+\Gamma _1 \left(2 \gamma  \left(9 \gamma ^2+8 \Omega^2\right)-\Gamma _1 \left(40 \gamma ^2+34 \Gamma _1 \gamma -17 \Omega ^2+8 \Gamma _1^2\right)\right)\right)}{\Gamma _1^2 \left(2 \gamma ^2+2\Gamma _1 \gamma +3 \Omega ^2\right)^5}.
%\eea
\end{widetext}

%\thispagestyle{empty}
%\cleardoublepage
\end {document}